\documentclass[aps,prd,twocolumn,prd,showpacs,showkeys,preprintnumbers,bibnotes,floatfix,longbibliography,notitlepage,nofootinbib,superscriptaddress,superscriptgaddress,amsmath
]{revtex4-2}
\include{jdefs}
\usepackage{graphicx}
\usepackage{placeins}
\usepackage{dcolumn}
\usepackage{bm}
\usepackage{tabularx}

\usepackage[colorlinks]{hyperref}
\hypersetup{
    pdfnewwindow=true,      
    colorlinks=true,       
    linkcolor=blue,          
    citecolor=blue,        
    filecolor=blue,      
    urlcolor=blue        
}

\usepackage{booktabs}
\usepackage{xcolor}
\usepackage{listings}
\usepackage{enumitem}
\usepackage{calligra}
\usepackage{xspace}

\newcommand{\srim}{\texttt{SRIM }}
\newcommand{\trim}{\texttt{TRIM }}

\newcommand{\sources}{\texttt{SOURCES-4A }}
\newcommand{\tendl}{\texttt{TENDL-2017 }}
\newcommand{\janis}{\texttt{JANIS4.O }}
\newcommand{\paleospec}{\texttt{paleoSpec }}

\begin{document}

\title{Refining the sensitivity of new physics searches with ancient minerals}

\author{Audrey Fung}
\email{audrey.fung@queensu.ca}
\affiliation{
Department of Physics, Engineering Physics and Astronomy,\\
Queen’s University, Kingston ON K7L 3N6, Canada
}
\affiliation{%
Arthur B. McDonald Canadian Astroparticle Physics Research Institute, Kingston ON K7L 3N6, Canada
}
\author{Thalles Lucas}
\email{thalles.lucas@queensu.ca}
\affiliation{
Department of Mechanical and Materials Engineering,\\
Queen’s University, Kingston ON K7L 3N6, Canada
}
\affiliation{%
Arthur B. McDonald Canadian Astroparticle Physics Research Institute, Kingston ON K7L 3N6, Canada
}
\author{Levente Balogh}
\email{levente.balogh@queensu.ca}
\affiliation{
Department of Mechanical and Materials Engineering,\\
Queen’s University, Kingston ON K7L 3N6, Canada
}
\affiliation{%
Arthur B. McDonald Canadian Astroparticle Physics Research Institute, Kingston ON K7L 3N6, Canada
}
\author{Matthew Leybourne}
\email{m.leybourne@queensu.ca}
\affiliation{
Department of Geological Sciences and Engineering,\\
Queen’s University, Kingston ON K7L 3N6, Canada
}
\affiliation{%
Arthur B. McDonald Canadian Astroparticle Physics Research Institute, Kingston ON K7L 3N6, Canada
}

\author{Aaron C. Vincent}
\email{aaron.vincent@queensu.ca}
\affiliation{
Department of Physics, Engineering Physics and Astronomy,\\
Queen’s University, Kingston ON K7L 3N6, Canada
}
\affiliation{%
Arthur B. McDonald Canadian Astroparticle Physics Research Institute, Kingston ON K7L 3N6, Canada
}
\affiliation{Laboratory for Particle Physics and Cosmology (LPPC), Harvard University,  Cambridge, MA 02138, USA}
\affiliation{Perimeter Institute for Theoretical Physics, Waterloo ON N2L 2Y5, Canada}

\date{\today}

\begin{abstract}
    
   Paleodetection has been proposed as a competitive method for detecting dark matter and other new physics interactions, complementing conventional direct detection experiments. In this work, we utilise \trim simulations to improve the modelling of track length distributions. Our findings suggest that previous studies have overestimated the number of tracks caused by weakly interacting particles, and that the lowest observable dark matter mass should be higher than previously predicted. These differences are mainly attributed to the fact that (a) the recoil energy-track length relation is not one-to-one, (b) at low recoil energies, a substantial fraction of recoils do not yield any tracks, and (c) at high energies, electronic stopping becomes dominant, resulting in a track length barrier at $\sim200$ nm. In addition to WIMPs, we also modelled tracks from generalised coherent elastic neutrino nucleus scattering (CE$\nu$NS) via new light mediators and estimated the projected sensitivity for these interactions. 
\end{abstract}

\maketitle

\section{Introduction}

There is overwhelming indirect evidence that dark matter exists, from galaxy rotation curves, to bullet cluster, to the CMB, but its microphysical properties remain unknown \cite{Planck:2018vyg}. Weakly interacting massive particles (WIMPs) have been, and remain, one of the most sought-after candidates. Thermally produced in the Early Universe, and generic to many extensions of the Standard Model (SM), such particles could have masses from $\sim$ GeV to TeV with weak-scale or lower cross sections. 
Currently, there are a number of direct detection experiments that are dedicated to detecting WIMPs, with the strongest limits above 10 GeV currently set by LZ \cite{LZCollaboration:2024lux}, and at lower masses by PICO-60 \cite{PICO:2019vsc}, DarkSide \cite{DarkSide-50:2022qzh}, NEWS-G \cite{NEWS-G:2024jms} and XENON-nT \cite{XENON:2023cxc}, depending on the (iso)spin structure of the DM-nucleon coupling. 
These direct detection experiments search for evidence of nuclear recoils in large detector volumes caused by the elastic scattering of dark matter particles on target material. These recoils can lead to scintillation, heating, ionization, or a combination of these, which can then be read out. Assuming backgrounds are well-understood, experiments are limited at low masses ($m_\chi$) by the threshold energy deposition required to see a signal, at low cross sections by detector volume and exposure time, and at high masses by the dark matter flux which scales as $m_\chi^{-1}$. 

It was realized early on that ancient minerals could provide a similar target material to search for heavy exotic particles \cite{Price:1986ky,Snowden-Ifft:1995zgn} As long as the nuclear recoil was energetic enough to displace multiple atoms, and therefore create a damage track in the material, the accumulated history of tracks could provide evidence---or constraints---on such particles. In fact, such searches are routinely used to estimate the age of some minerals using tracks produced after spontaneous uranium fission.
This line of research has recently been revived in the context of dark matter detection, both in recasting these early limits as dark matter constraints \cite{Acevedo:2021tbl}, as well as in the context of lower-mass WIMP-like dark matter \cite{Baum:2018tfw,Edwards:2018hcf,Drukier:2018pdy,Baum:2021chx,Baum:2021jak, Baum:2024eyr}. As the Earth traverses the dark matter halo in the Milky Way and dark matter scatters with nuclei within the crust, the resultant recoil could knock an atom off its lattice site. The first displaced atom (hereafter referred to as the primary knock-on atom, PKA)  moves across the lattice, leaving behind permanent damage tracks with lengths $x$ on nanometer scales, which can be identified using various microscopy techniques. In addition to dark matter detection, paleodetectors have been proposed to search for traces of neutrinos from nearby supernovae \cite{Baum:2019fqm}, diffuse supernova neutrino background (DSNB) \cite{Baum:2022wfc}, cosmic rays \cite{Galelli:2023tay, Caccianiga:2024otm, Mariani:2023mwb} and to characterize the time variation of the atmospheric \cite{Jordan:2020gxx} and solar \cite{Tapia-Arellano:2021cml} neutrino fluxes. There is also ongoing work on heavy dark matter in the form of composite dark matter, Q-balls, quark nuggets monopoles, etc. \cite{Baum:2023cct, Baum:2024eyr}.

As pointed out almost immediately \cite{Collar:1995aw}, paleodetectors potentially suffer from large backgrounds coming from cosmic rays, radioactive decays underground and neutrinos, and the signal from new physics is often expected to be smaller than the total background rate. A potential signal can therefore only be disentangled thanks to its expected track length spectrum $dR/dx$, in an analogous way to the deposited EM-equivalent energy spectrum reconstructed in neutrino detectors. Together, it means that a careful statistical treatment and an accurate modelling of the track length distribution are imperative. 

 Another, related challenge is that the track length due to a recoil event cannot be mapped directly to a single recoil energy $E_R$. Rather, a given $E_R$ can yield a range of different track lengths depending on the random walk the PKA takes inside the lattice, or in some cases, produce no track at all. As far as we can tell, all prior literature has assumed a one-to-one correspondence between $E_R$ and $x$ using the stopping length. Even the definition of the track length is subject to discussion, since real particle propagation in a lattice does not generally follow a straight line. This leads to overly optimistic projections, as the smearing due to $dE_R/dx$ reduces the overall signal height, but also leads to missed opportunities, as it erases important features that could be used to identify new signals. 

Finally, the choice of minerals used for paleodetection sensitivity analyses has often been cursory. A number of criteria have been identified for suitable paleodetection targets (see e.g. \cite{Drukier:2018pdy}):  1) be readily available from cores collected during deep drilling; 2) be able to record and preserve nuclear recoil damage as tracks; 3) have low molecular number density so that track length and energy resolution are optimized;
4) be amenable to imaging for track identification and quantification; 5) have been isolated at depths of ~ 5 km to 10 km; and 6) have maintained such depths for $>$ 500 Myr.  Previous studies have suggested several potential minerals that would satisfy at least some of these criteria, although not all minerals indicated satisfy all these criteria. Such candidate minerals include olivine, halite, sinjarite, and nchwaningite. Most of these minerals, other than olivine, are unsuitable as paleodetectors on geological, geochemical, and mechanical engineering grounds. Some of the minerals suggested are rare, and most are either hydrous or contain halogens; these characteristics mean that these minerals are soft, resulting in rapid annealing (healing) of any recoil damage, are readily altered by crustal fluids, and are mechanically unable to remain undeformed after burial to depths in the range of 5-10 km or more.

We have identified two candidate minerals that are likely to be robust to the issues mentioned above: olivine, (Fe,Mg)$_2$SiO$_4$, which has been previously studied in this context, and galena, PbS, which has not. Due to the computational requirements of our Monte Carlo simulations, we will  focus on olivine alone for this work. We do not expect results from galena to differ substantially, apart from some loss of sensitivity at low recoil energies due to the heavier target masses, and some enhancement in total rates from coherent scattering on lead. 

  In this work, we therefore endeavour to accurately predict the recoil spectra expected in ancient olivine. Where simple stopping power tables were previously used to obtain the $E_R$-$x$ correspondence, we will make use of the Monte Carlo software \trim (Transport of Ions in Matter \cite{Ziegler:2010bzy}) to obtain $dR/dx$ in a way that accounts for the full range of possible propagation scenarios for the PKA in the mineral, and define a track length that is consistent with the assumed read-out resolution and the inherently stochastic shape of particle tracks in minerals. 
  
  Such a treatment is necessary for both signal and background; we have also performed updated measurements of the uranium and thorium concentrations in samples of galena and olivine, which we will use as a benchmark to estimate the background neutron flux in this work.

  Combined, these considerations will lead to updated projections of the sensitivity of paleodetection experiments to dark matter. We will additionally examine the potential of new physics detection using solar neutrinos. Specifically, new light mediators   could result in increased interaction rates \cite{cerdeno2016physics,DeRomeri:2024dbv}, and therefore larger track rates from solar neutrinos than expected, with unique spectral features. 

In Sec. \ref{sec:theory}, we briefly present the standard approach to model the differential recoil rate $dR/dx$, and introduce a modification that incorporates additional track formation scenarios. In Secs. \ref{sec:wimp_theory}-\ref{sec:B-L_theory}, we review the computation of nuclear recoil spectra for WIMPs, and for neutrino-nuclear interactions via new light mediators and $B-L$ gauge bosons, respectively. In Sec. \ref{sec:backgrounds}, we outline the background sources considered in this work. In Sec. \ref{sec:trim}, we describe the implementation of \trim simulations and the characterisation of track length, followed by the track length distributions from new physics interactions in Sec. \ref{sec:drdx}. Finally, we present our results and constraints in Sec. \ref{sec:limits} and conclude in Sec. \ref{sec:conclusion}.

\section{theory}
\label{sec:theory}
\subsection{Integrated track length distribution}

Over the course of millions to billions of years underground, paleodetectors are expected to develop large densities of defect tracks, coming from fission, alpha and beta decays within the crust, atmospheric and extraterrestrial neutrinos, as well as possible new physics interactions. Dark matter-induced tracks cannot be identified on an event-wise basis; rather,  the track length distribution must be modelled in order to statistically distinguish signal from background. The track length distribution is usually modelled as 
\begin{equation}
\label{eq: old_drdx}
    \frac{dR}{dx}(x) = \sum_i \frac{dR_i}{dE_R}(E_R) \left\vert \frac{dE}{dx}(E_R) \right\vert_i,
\end{equation}
where $dR_i/dE_R$ is the recoil spectrum of element $i$ in the mineral and $dE/dx$ is the stopping power of element $i$. Stopping power is only a function of the target mineral, or specifically, the constituent elements and the structure of the mineral. It measures the expectation value of the differential energy loss of an irradiating particle per unit length of penetration into a material, rather than the exact trajectory-dependent values for individual ions. In previous studies, $\vert dE/dx 
\vert$ was obtained using stopping power tables from \srim \cite{Ziegler:2010bzy}, a software package that models energy loss of ions in a material and includes both nuclear and electronic stopping contributions.

However, in reality one expects a variety of stopping ranges due to the stochastic ion-ion interactions in the crystal lattice. Stopping power tables provide a quick way to estimate the expected track length from a recoil, but they assume a one-to-one relationship between the observed track length $x$ and recoil energy $E_R$:
\begin{equation}
\label{eq:x-E_sp}
    x_{avg}(E_R) = \int_0^{E_R} \left \vert\frac{dE}{dx}(E)\right \vert ^{-1} dE,
\end{equation}
where $x_{avg}$ represents the mean track length expected from a recoil energy $E_R$. In this work, we take into account the fact that one $E_R$ could yield a distribution of possible $x$ values, or conversely, a track length $x$ could result from a range of recoil energies $E_R$ with varying probabilities, $P(E_R\vert x)$. We model this using Monte Carlo data from \trim. In this case, the track length distribution is instead an integral over recoil energies:

\begin{equation}
    \label{eq:mydrdx}
    \frac{dR}{dx}(x) = \sum_i \int dE_R P_i(x\vert E_R) \frac{dR_i}{dE_R} (E_R)  \mathcal{P}_{i,track}(E_R).
\end{equation}
$\mathcal{P}_{i,track}(E_R)$ is the probability of track formation as a function of recoil energy, the details and implications of which will be discussed in Sec. \ref{sec:trim}.

\subsection{WIMP recoil spectrum}
\label{sec:wimp_theory}
Paleodetectors are generally sensitive to both spin-independent and spin-dependent interactions \cite{Drukier:2018pdy}. However, since the constituents of olivine almost all have zero nuclear spin, this mineral is far more sensitive to spin-independent interactions. We therefore focus on constant, spin-independent isoscalar DM-nucleus interactions. The expected nuclear recoil distribution from WIMP dark matter with isotope $i$ generally takes the form 
\begin{equation}
\label{eq:dRdE_general}
    \frac{dR_i}{dE_R} = \frac{\rho_\chi}{m_{N_i} m_\chi} \int dv v f(v) \frac{d\sigma_{\chi,{N_i}}}{dE_R}\ ,
\end{equation}
where $\rho_\chi = 0.3$ GeV cm$^{-3}$ \cite{Evans:2019PRD}, is the local density of dark matter, $m_{N_i}$ and $m_\chi$ are the masses of the target nucleus of interest and of WIMP particle respectively, $f(v)$ is the speed $v$ distribution of dark matter in the lab frame and $d\sigma_{\chi,N}/dE_R$ is the differential WIMP-nuclear interaction cross section. This quantity is a function of momentum transfer, but it can be written in terms of the cross-section at zero momentum transfer $\sigma_{\chi,{N_i}}(q=0)$
\begin{equation}
\label{eq:dSigdE}
    \frac{d \sigma_{\chi,{N_i}}}{dE_R} = \frac{m_{N_i}}{2 \mu_{\chi,{N_i}}^2v^2}\sigma_{\chi,{N_i}}(q=0) F_{i}^2(q)
\end{equation}
where $\mu_{\chi,{N_i}} = m_{N_i} m_\chi/(m_{N_i}+m_\chi)$ is the reduced mass of dark matter and nucleus, the nuclear form factor $F(q)$ is by definition the Fourier transform of the charge density profile of the nucleus, it takes into account the effect of the finite size of a nucleus with non-zero momentum exchange. Assuming the Helm density profile, the form factor can be expressed analytically \cite{Lewin:1995rx}:
\begin{align}
    F_i(qr_{n_i}) &= 3\frac{j_1(qr_n)}{qr_{n_i}}e^{-(qs)^2/2} \\
    &= \frac{3(\sin(qr_{n_i})-qr_{n_i}\cos(qr_{n_i}))}{(qr_{n_i})^3}e^{-(qs)^2/2} \nonumber
\end{align}
where $j_1$ is the first spherical Bessel function of the first kind. We took the best fit values for the effective nuclear radius $r_{n_i} = 1.14A^{1/3}$ fm and the nuclear skin thickness $s = 0.9$ fm  \cite{Lewin:1995rx}. 

The exact form of $\sigma_{\chi,N}(q=0)$ depends on specific particle physics models, but if we assume a generic scalar interaction, $\sigma_{\chi,N}(q=0)$ can be expressed as
\begin{align}
\label{eq: sigma0_simplified}
    \sigma_{\chi,{N_i}}(q=0) &= \frac{4\mu_{\chi, {N_i}}^2}{\pi}(Zf^p + (A-Z)f^n)^2 \\
    &\approx \frac{4\mu_{\chi, {N_i}}^2}{\pi}(Af^{p})^2 \nonumber
\end{align}
with $f^{p,n}$ being the \{proton or neutron\}-WIMP couplings, A and Z being the nucleon and proton numbers of the target nucleus respectively. The second line of Eq. \eqref{eq: sigma0_simplified} assumes $f^p \sim f^n$. Expressing WIMP recoil spectrum in terms of the spin-independent WIMP-nucleon cross section $\sigma_{SI}$ gives
\begin{equation}
    \frac{dR_i}{dE_R} = \frac{A_i^2 \sigma_{SI} F_i^2(E_R)}{2\mu_{\chi, {N_i}}^2 m_\chi} \rho_\chi \eta_\chi(v_{min}),
\end{equation}
where $\eta_\chi$ is defined as
\begin{align}
\label{eq:eta_x}
    \eta_\chi  &\equiv \int_{v_{min}} d^3\mathbf{v} \frac{f(\mathbf{v})}{v} \\
    &= \int_{v_{min}}^{v_{esc}} dv v^2 2\pi \frac{f(v)}{v} \nonumber
\end{align}
The integral is kinematically bounded from below with the minimum incoming velocity required to produce a recoil of energy $E_R$, $v_{min} = \sqrt{m_{N_i}E_R/2\mu_{\chi, {N_i}}^2}$. We used the truncated Maxwellian velocity distributions for $f(v)$ and the corresponding astrophysical parameters given in Ref. \cite{Evans:2019PRD}:
\begin{equation}
    f(\mathbf{v}) = \frac{1}{(2\pi \sigma_v^2)^{3/2}N} \exp\left(-\frac{\vert \mathbf{v} \vert^2}{2\sigma_v^2}\right) \Theta(v_{esc}-\vert \mathbf{v}\vert)
\end{equation}
which $f(\mathbf{v})$ is replaced by $f(v)$ in Eq. \eqref{eq:eta_x} due to isotropy. The Heaviside step function $\Theta$ ensures that velocities greater than the escape velocity of the Milky Way do not contribute, implying a hard cutoff in the energy spectra at the corresponding recoil energy. The normalization factor $N$ is introduced to account for the truncation: 
\begin{equation}
    N = \text{erf} \left(\frac{v_{esc}}{\sqrt{2}\sigma_v}\right) - \sqrt{\frac{2}{\pi}}\frac{v_{esc}}{\sigma_v}\exp \left(-\frac{v_{esc}^2}{2\sigma_v^2}\right). 
\end{equation}

We took $\sigma_v  = (223/\sqrt{2})$ km s$^{-1}$ and $v_{esc} = 544$ km s$^{-1}$ from Table 1 of Ref. \cite{Evans:2019PRD}. These values are derived from a combination of empirical observations and hydrodynamic simulations, and do not account for e.g. dark matter substructure, which could enhance or suppress rates \cite{DEAP:2020iwi,Smith-Orlik:2023kyl}.

\subsection{Coherent elastic neutrino-nucleus scattering (CE$\nu$NS) via new light mediators} 
\label{sec:cevens_theory} 

At low energies, neutrinos undergo neutral current interactions with an entire nucleus, scattering coherently with a cross section that scales approximately with the square of the nucleon number. This enhancement enables a more precise measurement of neutrino properties and provides a window into potential new neutrino interactions, including those mediated by scalar, vectors and axial-vectors. 
Solar neutrinos are expected to be one of the major sources of backgrounds in dark matter direct detection, they contribute significantly to the so-called neutrino fog. Since solar neutrinos are abundant and have energies in the keV range, where CE$\nu$NS is relevant, instead of treating this as a limitation, we turn this to our advantage and look for non-standard neutrino-nuclear interactions. 

The differential nuclear recoil rate for a neutrino per unit target mass can be written as 
\begin{equation}
    \frac{dR}{dE_R} = \frac{1}{m_T}\int dE_\nu \frac{d\Phi}{dE_\nu}\frac{d\sigma_{\nu,N}}{dE_R}
\end{equation}
where $d\Phi/dE_\nu$ is the differential neutrino flux at Earth per unit neutrino  energy $E_\nu$, $d\sigma_{\nu,N}/dE_R$ is the differential cross section per unit recoil energy $E_R$ of a nucleus. We use the solar neutrino fluxes provided in Refs. \cite{Gonzalo:2023mdh,Gonzalez-Garcia:2023kva}. For standard model interactions, the coherent differential cross section reads 
\begin{equation}
    \frac{d\sigma_{\nu,N}}{dE_R} = \frac{G_F^2}{4\pi}Q_v^2 m_NF^2(q)\left(1-\frac{E_R}{E_{R,max}(E_\nu)}\right)
\end{equation}
where $E_{R,max} = 2E_\nu^2/(m_N+2E_\nu)$ denotes the maximum recoil energy  at a given $E_\nu$. $G_F$ is the Fermi constant, $Q_v$ is the coherence factor parametrised by proton number $Z$, neutron number $(A-Z)$ and weak mixing angle $\theta_W$ as
\begin{equation}
    Q_v \equiv (A-Z) - (1-4\sin^2\theta_W)Z.
\end{equation}
At low recoil energies relevant for solar neutrino scattering, $\sin^2\theta_W = 0.023867$ \cite{erler2005weak}.

CE$\nu$NS is known to be sensitive to new physics, which has been studied extensively in Refs. 
\cite{cerdeno2016physics,barranco2005probing,dutta2016sensitivity,kosmas2015standard,scholberg2006prospects} . We focus on the model-independent interactions via new light mediators as studied in Ref. \cite{cerdeno2016physics}, which focused on scalar $\phi$, vector $Z'$ and axial-vector $a$ mediators. In the presence of such new force carriers, the neutrino-nucleus cross sections are modified by adding extra contributions to the standard model term $(d\sigma/dE_R)_{SM}$\cite{cerdeno2016physics}:
\begin{align}
\label{eq:dsigdE_scalar}
    \left(\frac{d\sigma}{dE_R}\right)_{\phi} & = \left(\frac{d\sigma}{dE_R}\right)_{SM} \\
    & + \frac{Q_s^{'2}m_N^2E_R}{4\pi E_\nu^2(2E_R m_N+m_\phi)^2} \nonumber,
\end{align}

\begin{align}
\label{eq:dsigdE_vector}
    \left(\frac{d\sigma}{dE_R}\right)_{Z'} & = \left(\frac{d\sigma}{dE_R}\right)_{SM} \\
    & -\frac{G_Fm_NQ_vQ_v^{'2}(2E_\nu^2-E_Rm_N)}{2\sqrt{2}\pi E_\nu^2(2E_Rm_N+m_{Z'}^2)} \nonumber \\
    & +\frac{Q_v^{'2}m_N(2E_\nu ^2-E_Rm_N)}{4\pi E_\nu^2(2E_Rm_N+m_{Z'}^2)^2}, \nonumber
\end{align}

\begin{align}
\label{eq:dsigdE_axialVec}
    \left(\frac{d\sigma}{dE_R}\right)_{a} & = \left(\frac{d\sigma}{dE_R}\right)_{SM} \\
    & +\frac{G_Fm_NQ_aQ_a^{'2}(2E_\nu^2+E_Rm_N)}{2\sqrt{2}\pi E_\nu^2(2E_Rm_N+m_{a}^2)} \nonumber \\
    & -\frac{G_Fm_NQ_vQ'_aE_\nu E_R}{2\sqrt{2}\pi E_\nu^2(2E_Rm_N+m_{a}^2)} \nonumber \\
    & +\frac{Q_a^{'2}m_N(2E_\nu^2+E_Rm_N)}{4\pi E_\nu^2(2E_Rm_N+m_{a}^2)^2}. \nonumber 
\end{align}
with
\begin{equation}
    Q_a \approx 1.3 S_N
\end{equation}
\cite{cerdeno2016physics} and 
and the new coherent factors defined as
\begin{align}
    & Q_s' \approx (14A + 1.1Z)\ g_{\nu, \phi} g_{q, \phi} \, ,\\
    & Q_v' = 3A \ g_{\nu, Z'} g_{q, Z'}\, , \\
    & Q_a' \approx  0.3S_N \ g_{\nu, a} g_{q, a}\, ,
\end{align}
where $S_N$ is nuclear spin and $g_{\nu, X}$ and $g_{q, X}$ are the couplings between neutrino with mediators $X = {\phi, Z', a}$, and quarks with mediators respectively. We assume the interactions are flavour-independent. We also define:
\begin{align}
    & g_\phi = \sqrt{g_{\nu, \phi} g_{q, \phi}}\, , \\
    & g_V = \sqrt{g_{\nu, Z'} g_{q, Z'}} \, , \\
    & g_a = \sqrt{g_{\nu, a} g_{q, a}}\, .
\end{align}
so that $Q_s'$, $Q_v'$ and $Q_a'$ can now be expressed in terms of a single effective coupling $g_\phi^2$, $g_V^2$ and $g_a^2$ respectively.

\subsection{$U(1)_{B-L}$}
\label{sec:B-L_theory}
 The CE$\nu$NS cross sections for vectors can also be extended to include light $U(1)_{B-L}$ vector gauge boson. However, since neutrino carries a $B-L$ charge of $q_{B-L}^{\nu} = -1$ while $q_{B-L}^{quarks} = 1/3$ \cite{Okada:2018ktp}, the coherent factor becomes 
\begin{equation}
    Q_{B-L}' = 3A(-\frac{1}{3} g_{B-L}^2) = -\frac{1}{3}Q_v'
\end{equation}
where $-1/3$ comes from the fact that  $Q_{B-L}^{quarks} = -1/3Q_{B-L}^{\nu}$, and $g_{B-L}$ is the $B-L$ gauge boson coupling. The cross section for $B-L$ is then
\begin{align}
\label{eq:dsigdE_BL}
    \left(\frac{d\sigma}{dE_R}\right)_{B-L} & = \left(\frac{d\sigma}{dE_R}\right)_{SM} \\
    & +\frac{G_Fm_NQ_vQ_v^{'2}(2E_\nu^2-E_Rm_N)}{(3)2\sqrt{2}\pi E_\nu^2(2E_Rm_N+m_{Z'}^2)} \nonumber \\
    & +\frac{Q_v^{'2}m_N(2E_\nu ^2-E_Rm_N)}{(9)4\pi E_\nu^2(2E_Rm_N+m_{Z'}^2)^2}. \nonumber
\end{align}
The interference term is positive here, and we thus expect that $B-L$ mediators would always increase the interaction rate, in contrast to the case for vector mediators. 

\section{Backgrounds}
\label{sec:backgrounds}
Backgrounds are expected to be dominant in paleodetection. Refs. \cite{Drukier:2018pdy, Baum:2021jak} have performed detailed studies on the possible sources and provides relevant background spectra and fluxes, some of which we adopt in our modelling. Below, we review all the background sources considered in this study. The expected errors we used for each of these background sources are summarised in Table \ref{tab:nuisance}.

\begin{figure}
    \centering
    \includegraphics[width=\linewidth]{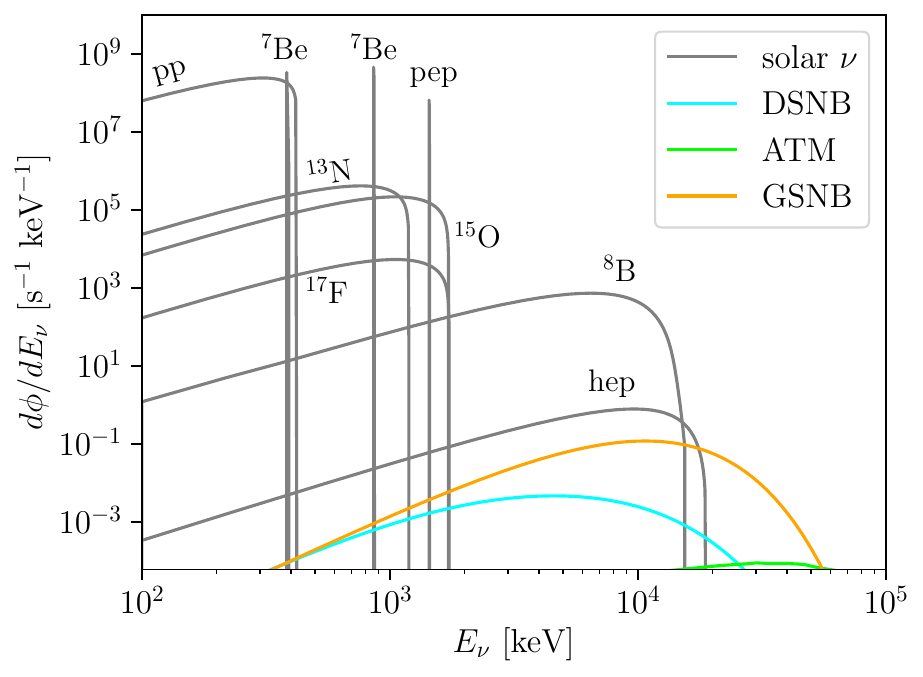}
    \caption{Spectra of neutrinos expected in paleodetectors. The grey curves show the solar neutrino contributions from different fusion processes. In cyan, lime and orange are the diffuse supernova neutrino background, atmospheric neutrinos from cosmic rays and core-collapse supernova neutrinos from the Milky Way respectively.}
    \label{fig:nu_fluxes}
\end{figure}

\begin{figure}
    \centering
    \includegraphics[width=\linewidth]{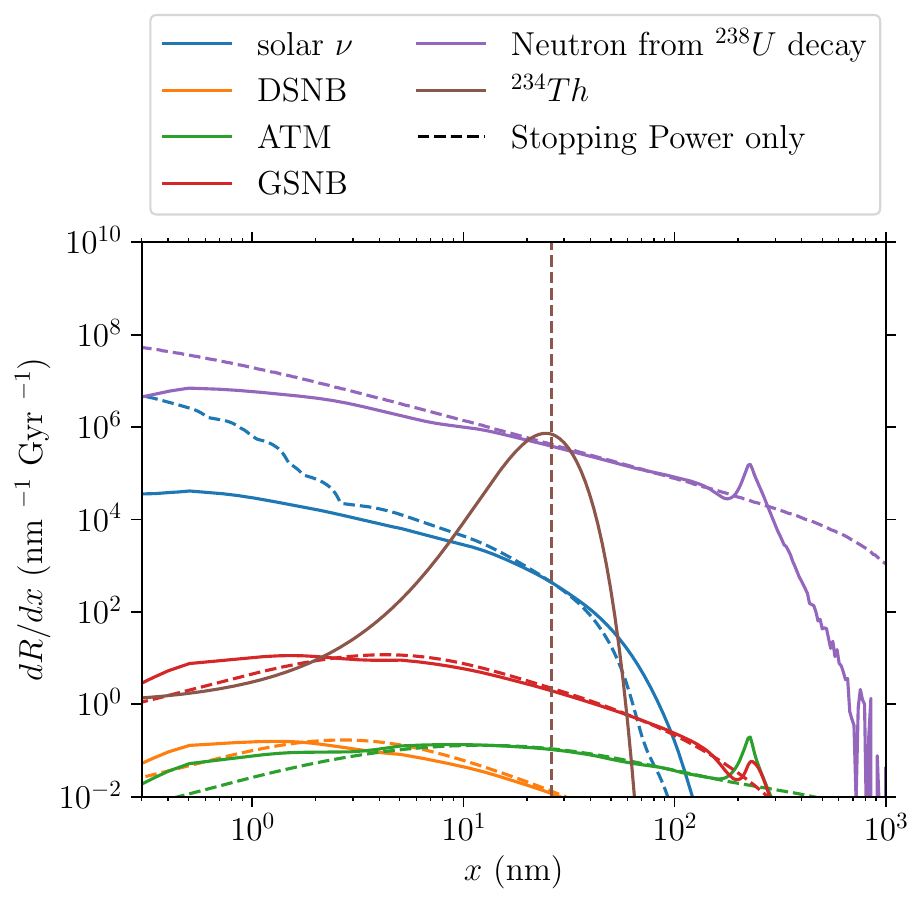}
    \caption{Differential recoil rates from backgrounds as a function of track length. We assume a benchmark value of $10^{-4}$ ppm (or $10^{-10}$g/g) for $^{238}$U concentration in the calculation of neutron fluxes.}
    \label{fig:bkg_tradks}
\end{figure}

\paragraph{Solar Neutrinos:} While Solar neutrinos serve as the detection signals for generalised neutrinos interactions, they also constitute the dominant irreducible neutrino background for dark matter detection. Figure \ref{fig:nu_fluxes} shows the solar neutrino spectra, which were obtained from Ref. \cite{Gonzalo:2023mdh}, with their normalisations taken from Ref. \cite{Gonzalez-Garcia:2023kva}. Figure \ref{fig:bkg_tradks} shows that solar neutrinos primarily contribute to tracks with $x<100$ nm, which is where most of our signals of interest lie, as we will see in Sec. \ref{sec:drdx}. As reported e.g. in Ref. \cite{Serenelli:2016nms}, the uncertainties for the flux normalisations from different fusion processes range from 0.6\% to 17\%. For simplicity, we adopt a realistic value of 14\% for the total flux normalisation, which corresponds to the theory error on the $^8$B flux that is dominant in the signal region. 
\newline
\paragraph{Supernova Neutrinos:} Neutrinos (and antineutrinos) emitted from all core-collapse supernovae across all times in the observable universe form the Diffuse Supernova Neutrino Background (DSNB). It primarily spans the energy range of $10^{3} - 10^{4}$ keV and is expected to be isotropic and time-independent. While the DSNB has not yet been detected, Ref. \cite{Baum:2019fqm} showed that it could leave tracks ranging from $x =1 - 100$ nm in minerals. They have also shown that supernova neutrinos from the Milky Way, which they referred to as the Galactic Supernova Neutrino background (GSNB), produce an non-negligible number of events. We obtained the DSNB and GSNB fluxes along with their theoretical uncertainties (put conservatively at 100 \%) from Ref. \cite{Baum:2019fqm}.
\newline
\paragraph{Atmospheric Neutrinos:}Cosmic rays hitting the Earth's atmosphere also produce energetic neutrinos through the production and decay of pions and kaons, with a measurable flux primarily in the $10^4 - 10^5$ keV range, as seen faintly emerging from the bottom of Figure \ref{fig:nu_fluxes}. Despite being subdominant, atmospheric neutrinos are still expected to leave tracks beyond $100$nm. We took atmospheric neutrino fluxes from Ref.~\cite{Baum:2021jak}, which were adapted from Ref.~\cite{OHare:2020lva}. Although estimates exist for the measurement errors in the flux normalisation, the complexities of hadron production mechanisms coupled with the cosmic ray spectrum introduce additional uncertainties. To be conservative we assume a 100\% error.
\newline
\paragraph{Neutrons from $^{238}$U decay chains:} Radioactive $^{238}U$ can undergo $\alpha$-decay and spontaneous fission, resulting in energetic neutrons which can span several orders of magnitude in track lengths and constituting the dominant background in paleodetection. 
The number of neutrons produced directly scales with the concentration of $^{238}U$. Upon measuring the $^{238}U$ concentration of eight Olivine samples, we observed a peak around $10^{-3.4}$ ppm ($10^{-9.4}$ grams of $^{238}U$ per gram of target mineral). However, the detection threshold in this work was also at $10^{-3.4}$ ppm. This suggests that the true distribution likely peaks below the threshold. For the purpose of this work, we used $10^{-4}$ ppm ($10^{-10}$ g/g) as a benchmark value and assume a 1\% error, which is commonly achievable with the use of inductively coupled plasma mass spectrometry (ICP-MS) and should be possible in a dedicated study.

To model the spectral shape of nuclear recoils expected from neutrons, we used the tabulated values for the constituents of Olivine from \paleospec \cite{Baum:2018tfw, Drukier:2018pdy, Baum:2019fqm, Baum:2021jak}. In their work, they computed the neutron spectra from $\alpha$-decay and spontaneous fission using \sources \cite{WilsonW.B.2009SAcf} and obtained the recoil spectra for various target minerals from \tendl \cite{Koning:2012zqy, Plompen:2017lvx,Sublet:2015,Fleming:2015} and \janis \cite{SopperaN.2014J4AI}. 
\newline
\paragraph{Thorium-234 from radioactive decay}
$^{238}$U undergoes radioactive decay, producing an alpha particle and a $^{234}$Th nucleus in the first alpha decay of the chain. Since this is a two-body decay, $^{234}$Th receives a monoenergetic recoil energy of 72 keV. $^{234}$Th is the only source energetic enough to create observable tracks directly, rather than relying on recoiled PKAs in the minerals to produce tracks. Ref. \cite{Drukier:2018pdy} estimates the number density of $^{234}$Th to be approximately $10^6$ g$^{-1}$ for $^{238}U$ concentration at 0.01 ppb ($10^{-11}$ g/g), which would be $10^7$ g$^{-1}$ with our benchmark concentration at $0.1$ ppb ($10^{-10}$ g/g). 
\newline
\paragraph{Intrinsic Crystal Defects:} Minerals often contain impurities that cause intrinsic lattice defects to form upon crystallisation. These defects can be point-like (0D), linear (1D), planar (2D), or volumetric (3D). In particular, linear dislocations resemble nuclear recoil tracks. However, dislocation typically form loops or terminate at other defects, such as grain boundaries, where they extend across an entire grain \cite{DemouchySylvie2021Dio}, which is at least of order $\mathcal{O}(10^3)$ nm. These tracks should be distinguishable from $\mathcal{O}(1) - \mathcal{O}(100)$ nm tracks caused by nuclear recoils. 

\begin{table}[h]
\caption{Experimental uncertainties of nuisance parameters that are profiled over in our projections; see main text for details.}
 \label{tab:nuisance}
\centering
\begin{tabularx}{\linewidth}{@{} X r @{}}
\toprule
\textbf{Parameter $n_i$} & \textbf{Error $\sigma_i$} \\
\midrule
$\nu_{\text{solar}}$ flux        & 14\%     \\
$\nu_{\text{DSNB}}$ flux         & 100\%    \\
$\nu_{\text{GSNB}}$ flux         & 100\%    \\
$\nu_{\text{ATM}}$ flux          & 100\%    \\
$^{238}$U concentration          & 1\%      \\
Mineral age $t$                  & 5\%      \\
Mineral mass $m_{\text{olivine}}$ & 0.01\%   \\
\bottomrule
\end{tabularx}
\end{table}

\section{Monte Carlo Simulations}
\subsection{Track Length Characterisation}
\label{sec:trim}
We employ \trim, a Monte Carlo simulation program within \srim that calculates the transport of traversing ions within an amorphous target, and their interactions with constituent atoms. These interactions are evaluated using binary collision approximation and the free-flight path approximation \cite{Ziegler:2010bzy}. \trim uses the Kinchin-Pease model in its quick damage calculation mode to estimate the number of displaced atoms based on the energy transferred by the PKA, and records the coordinates of each resulting displacement. This allows for a fast approximation without simulating the full cascade of secondary recoils. A track length can intuitively be taken as arc length spanned by the resulting vacant lattice sites. The dots in Figure \ref{fig:bezier_track} show an example of a series of vacant sites produced by a PKA with recoil energy $E_R$ at 90 keV. Ions hit by the PKA also create a cascade of secondary recoils, which lead to a broadening of the track. For computational tractability, we model the track as the sequence of PKA interaction sites. 

\begin{figure*}[t]
    \centering
    \includegraphics[width=\textwidth]{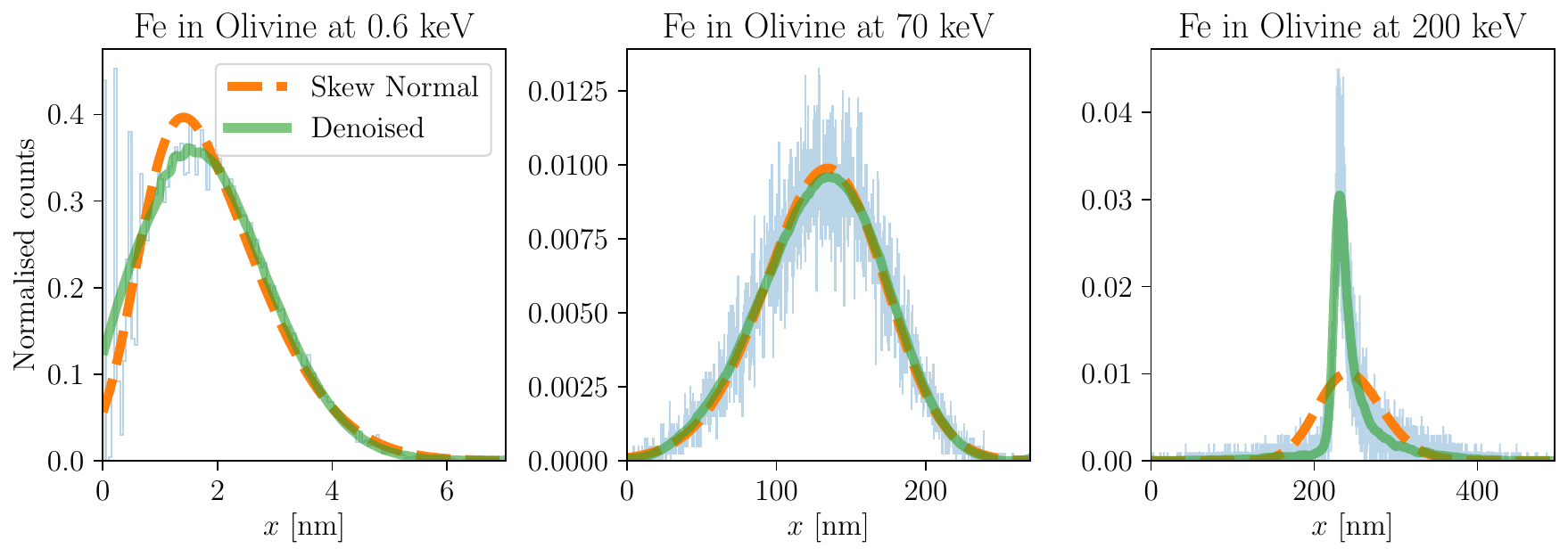}
    \caption{Histograms of track length distributions for primary knock-on atom recoil energies of $0.6$ keV, $70$ keV, and $200$ keV respectively. Overlaid are the fitted distribution functions by using a skew normal distribution (dashed) and by direct denoising of the data (solid). }
    \label{fig:fitting_dist}
\end{figure*}

We simulated scenarios in which external particles knock a PKA off its lattice site in olivine, imparting a recoil energy $E_R$. The PKA subsequently interacts with other ions in the medium, and by running  20,000 individual simulations (40,000 for $E_R \leq 1$ keV), we generated a statistical distribution of possible track lengths. This process was repeated for 35 values of $E_R$ ranging from 0.1 keV to 1000 keV with sampling adapted to regions of rapid change. This yielded an array of 35 normalized fixed-energy track length distributions. These distributions were then combined to construct the quantity  $P(x\vert E_R)$ in Eq. \eqref{eq:mydrdx}. Figure \ref{fig:fitting_dist} shows three examples of track length distributions for recoil energies $E_R$ of 0.6 keV, 70 keV, and 200 keV, respectively.

Before analysing the simulation results, we first discuss how track length is modelled based on the locations of vacant sites  produced by PKAs. The most direct approach is to define track length as the arc length spanned by the vacant lattice sites. However, in reality, secondary cascades of PKA cause multiple collision sites to appear as the same part of the track. Since we did not track any secondary cascades, and to avoid overestimating the track length, we modelled the track length using a Bézier parameterisation. We selected the lowest degree $n$ of correction necessary to achieve an error just below the considered readout resolution. Taking the track in Figure \ref{fig:bezier_track} as an example, $n = 2$ is the optimal fit as it is the minimum $n$ that brings the error just below $10$ nm, whereas $n=3$ would result in over-fitting, and thus overestimate the visible track length. The details of this implementation are provided in Appendix \ref{sec: bezier}.  

\begin{figure}
    \centering
    \includegraphics[width=0.51\textwidth]{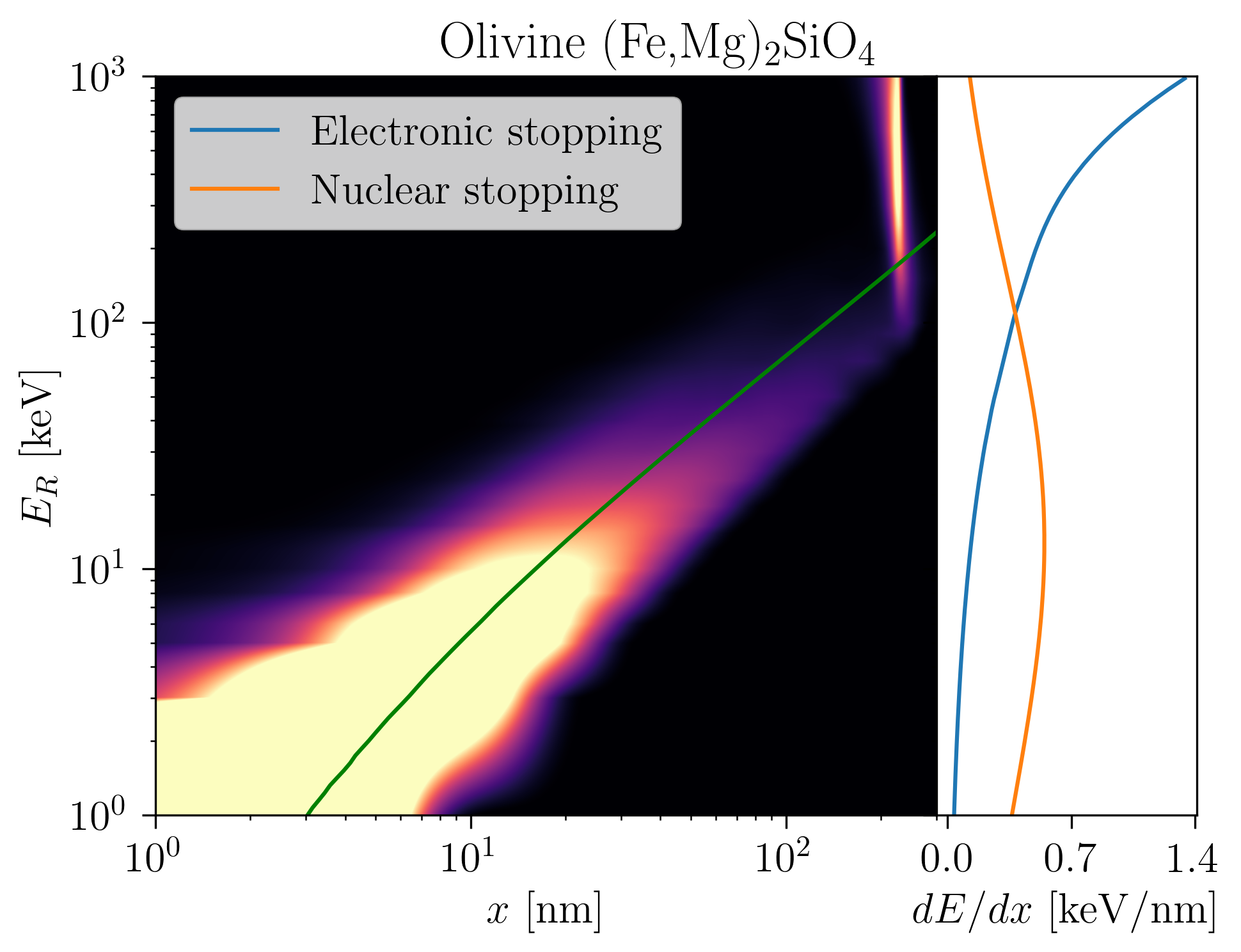}
    \caption{A 2D density plot where each horizontal slice represents a 1D normalised track length distribution at a given $E_R$ in olivine, with lighter colours indicating higher density. Note that this is not a joint probability distribution; hence its two-dimensional integral is not normalised. Overlaid in green is the $E_R$-$x$ relation obtained from stopping power tables. Plotted on the right is the nuclear stopping power (blue) and electronic stopping power (orange) as a function of recoil energy $E_R$.}
    \label{fig: 2d_track_density}
\end{figure}

To achieve smooth interpolation of track length distributions between energy points, the raw distributions are fitted with smooth functions to minimize interpolation noise. For $E_R < 40$ keV, skew normal distributions provide a reasonable fit; however, for $E_R \geq 40$ keV, even the best-fit parameters fail to capture the characteristic spiky profile observed at high recoil energies. In such cases, we denoise the distributions using Scipy signal processing package \texttt{scipy.signal} to obtain smooth profiles. Figure \ref{fig: 2d_track_density} shows a density plot of a 2D function $f(x,E_R)$, illustrating how the track length distributions $P(x\vert E_R)$ vary with $E_R$. Key features of this plot will be explained in the following paragraphs:

\begin{figure}
    \centering
    \includegraphics[width=\linewidth]{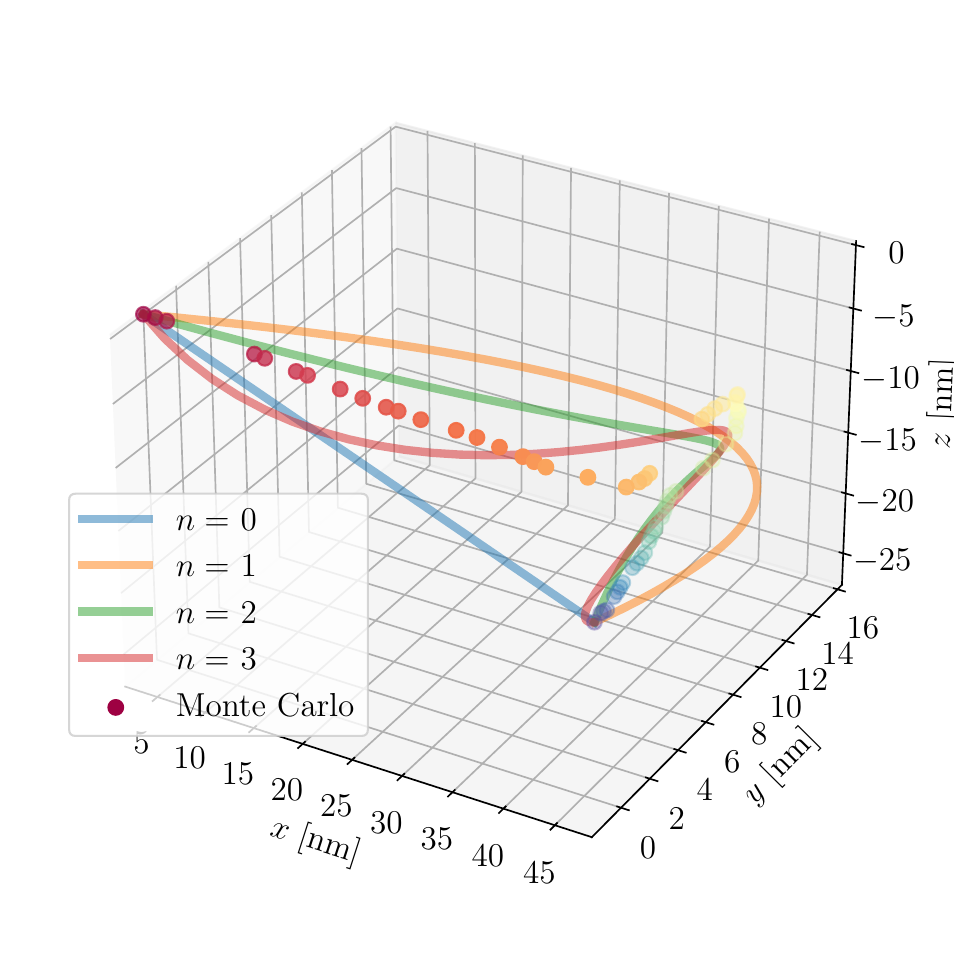}
    \caption{An example of collision sites of a Si PKA with $E_R = 90$ keV, represented by coloured dots, with red indicating earlier times and blue indicating later times. Bézier curves of degrees $n = 0$, $1$, $2$ and $3$ are also shown. }
    \label{fig:bezier_track}
\end{figure}

\begin{figure}
    \centering
    \includegraphics[width=\linewidth]{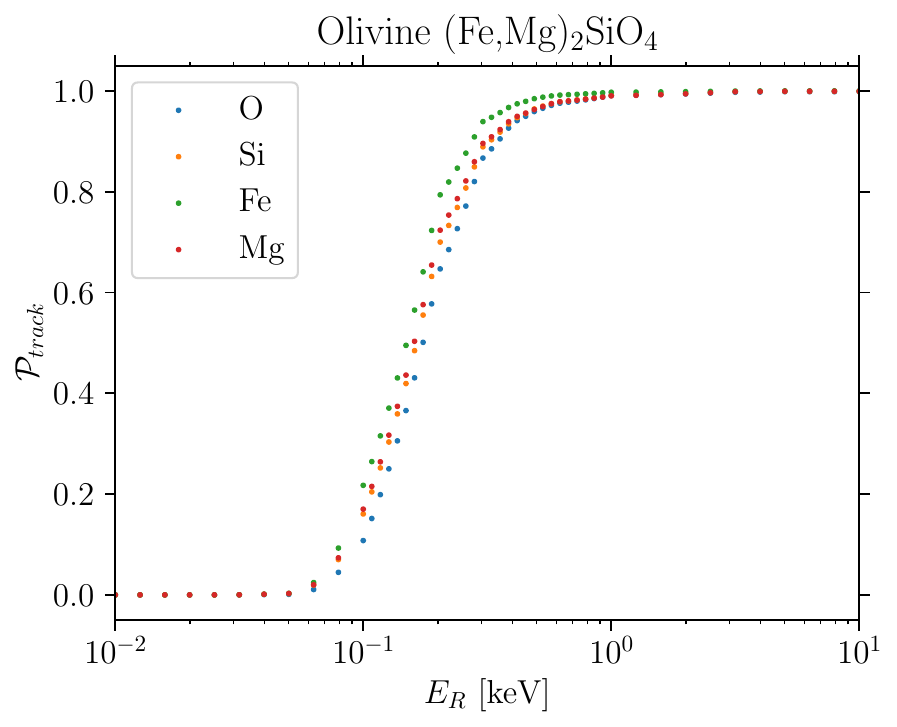}
    \caption{Probability of track formation in olivine as a function of PKA energy from TRIM simulations.}
    \label{fig:Prob_non0Track}
\end{figure}

\begin{enumerate}
    \item As mentioned, $P(x\vert E_R)$ deviates from a Gaussian-like profile for $E_R > 40$ keV as shown in Figure \ref{fig:fitting_dist}, indicating that distributions obtained from stopping power alone, even with Gaussian convolution, might not accurately represent the actual distribution especially at high energies. Moreover, the observed spread is intrinsic and thus two sources of broadening must be considered: one from the inherent dispersion of track lengths and another from the finite resolution of readout methods.
    
    \item Since at least two collision points are needed to define the endpoints of a track, any recoil that results in one or fewer collisions will have a track length of zero. We interpret this as these collisions not resulting in any track. Upon simulating 40,000 PKA interactions for each $E_R < 5$ keV, we found that as recoil energy decreases, the fraction of event resulting in non-zero track length decreases as expected, as shown in Figure \ref{fig:Prob_non0Track}. We define the probability $\mathcal{P}_{track}$ that a track is produced given a recoil as the fraction of tracks with nonzero length. Figure \ref{fig:Prob_non0Track} is similar for each element and is well fitted by a hyperbolic tangent function: 
    \begin{equation}
    \label{eq:Ptrack}
        \mathcal{P}_{track} = \frac{1}{2}\tanh(a\log_{10} E_R + b) + \frac{1}{2}, 
    \end{equation} 
    where $a$ and $b$ are constants listed in table \ref{tab:P_track}. This effect was not accounted for in Eq.~\eqref{eq: old_drdx}, where the total number of tracks is assumed to equal the total number of recoils, as described by the recoil spectrum of the incoming particle species of interest. To address this, we multiply the recoil rate by the probability of track formation $\mathcal{P}_{track}(E_R)$ in Eq. \eqref{eq:mydrdx}. The probability reaches 1 for tracks $x \gtrsim 1$ nm, meaning this adjustment does not affect low-resolution readouts. However, for high-resolution readouts, such as the $1$ nm resolution considered in this study, tracks at the measurement threshold of $0.5$ nm are predicted to occur at only half the rate estimated in previous studies using stopping power only. As we will see in Sec. \ref{sec:limits}, this leads to a decrease in sensitivity at low masses.
    
    \item Figure \ref{fig: 2d_track_density} presents the normalised track length distributions as a function of both $E_R$ and track length $x$. The track length-recoil energy relationship obtained from stopping power tables generally follows the modes of the distributions. However, at higher energies, an obvious deviation is observed: a plateau appears at $x \approx 200$ nm for $E_R \gtrsim 200$ keV, indicating a pile-up around this track length. This can be attributed to electronic stopping, such as ionization, dominating over nuclear stopping at high recoil energies, resulting in shorter tracks despite more energetic recoils. This effect is evident on the right side of Figure \ref{fig: 2d_track_density}, where the plateau appears when electronic stopping becomes dominant. 

    \item As mentioned in Sec. 
    \ref{sec:wimp_theory}, there is a maximum recoil energy corresponding to the escape velocity of the Milk Way. With the one-to-one track length assumption, a hard cutoff at the corresponding track length is therefore expected. However, with  realistic track length distribution, a significant amount of tracks longer than the cutoffs are observed.
\end{enumerate}

\begin{table}[h]
\begin{center}
\caption{Best fit constants $a$ and $b$ for Eq.~\eqref{eq:Ptrack} for the constituent elements of olivine.}
\label{tab:P_track}
\begin{tabularx}{\linewidth}{ X X r } 
\toprule
 Element & $a$ & $b$ \\  
 \midrule
 O &    3.867 & 2.917 \\ 
 Si & 3.780 & 2.972 \\ 
 Fe & 4.146 & 3.453 \\
 Mg & 3.830 & 3.050 \\
 \bottomrule
 
\end{tabularx}
\end{center}
\end{table}



\begin{figure*}
    \centering
        \includegraphics[width=0.48\textwidth]{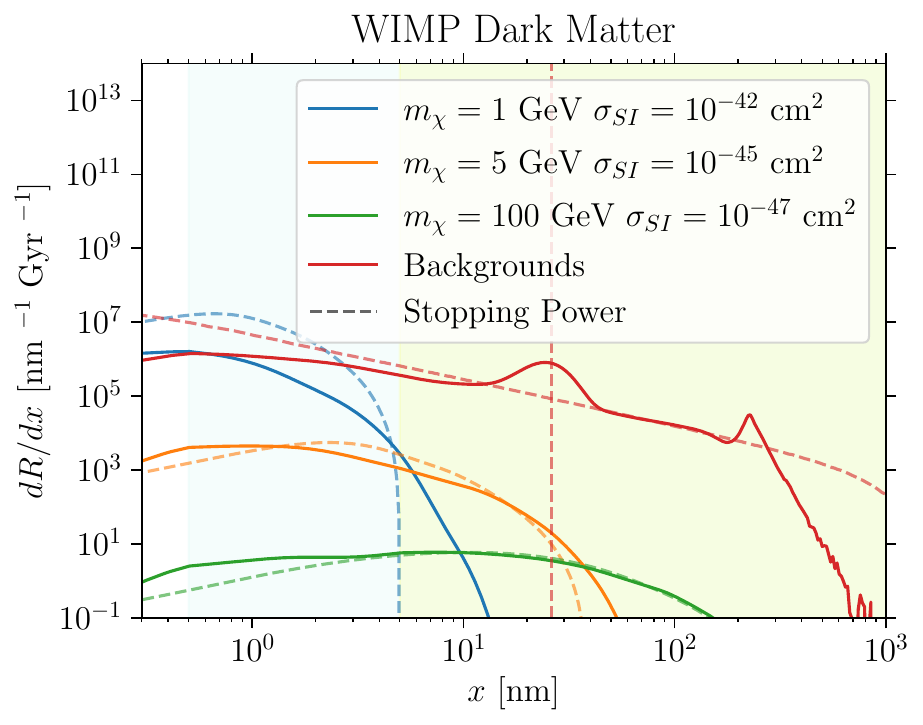}\quad
    \includegraphics[width=0.48\textwidth]{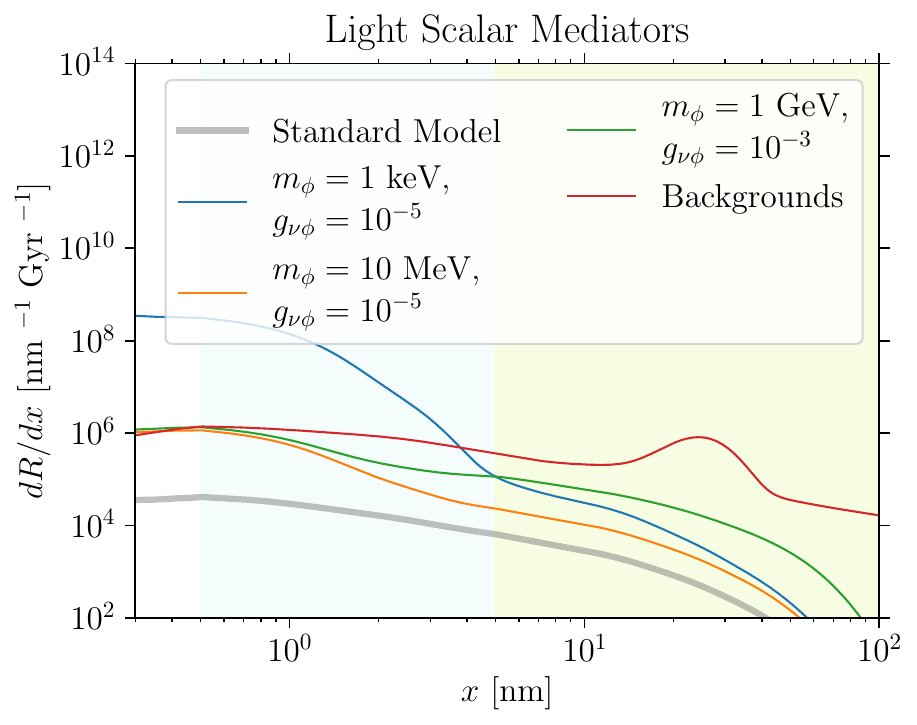}\\
    \includegraphics[width=0.48\textwidth]{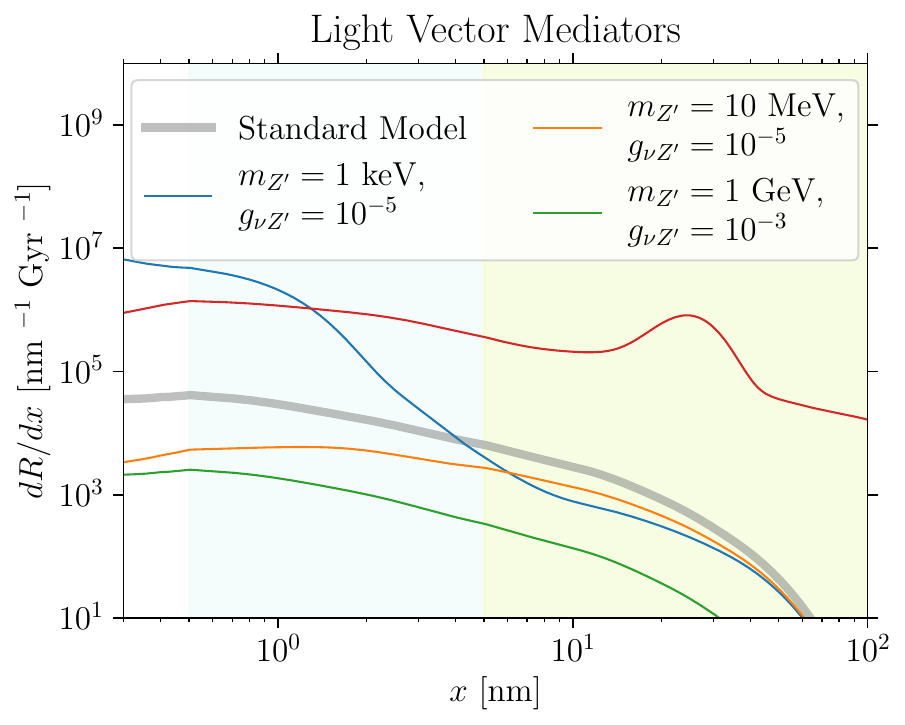}\quad
    \includegraphics[width=0.48\textwidth]{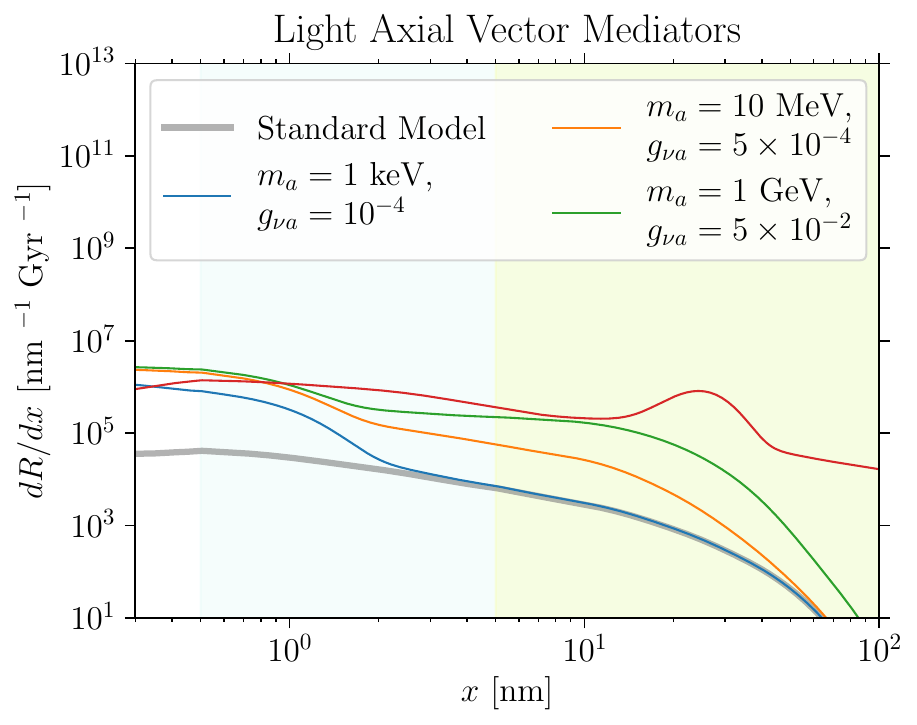}
    \caption{Differential recoil rate of (top-left) WIMPs, and solar neutrinos with new (top-right) light scalar mediators, (bottom-left) light vector mediators and (bottom-right) light axial-vector mediators as a function of track length. The solid lines represent results incorporating TRIM modelling, while the dashed lines correspond to stopping-power-only calculations. The cyan-shaded region indicates the track length range detectable with high-resolution readout, whereas the lime-shaded region highlights the range for low-resolution readout.}
    \label{fig:dRdx_sp_comparison}
\end{figure*}

\subsection{Track Length Distributions from New Physics Interactions in Olivine}
\label{sec:drdx}

\subsubsection{WIMPs}
The top-left panel of Figure \ref{fig:dRdx_sp_comparison} shows the differential recoil rate as a function of track length for three different combinations of dark matter masses and cross section: $m_\chi = 1$ GeV, $\sigma_{SI} = 10^{-47}$ cm$^2$; $m_\chi = 5$ GeV, $\sigma_{SI} = 10^{-45}$ cm$^2$; and $m_\chi = 100$ GeV, $\sigma_{SI} = 10^{-42}$ cm$^2$. The green shaded region highlights tracks that would be visible in our ``low resolution'' scenario ($x > 10/2$ nm), while the blue shading highlights the gain from a high-resolution (1/2 nm) readout. Solid lines show results using our \trim simulations, while dotted lines are equivalent spectra computed using stopping power only as done in previous works. 

The track length distribution for dark matter with $m_\chi = 1$ GeV generally comes from less energetic recoils, thus shorter tracks. In contract, heavier dark matter $m_\chi = 100$ GeV leads to more energetic but less frequent recoils due to a smaller DM flux. In most cases, using a realistic track length distribution tends to flatten the distribution. As mentioned in point \# 4 of Sec. \ref{sec:trim},  longer tracks than expected (e.g. $x \approx 5$ nm for $m_\chi = 1$ GeV) can now be observed. 

However, for the neutron background, we observe the opposite trend: our calculations predict a lower rate than the stopping power-only case. This is consistent with our prediction that recoils $\gtrsim 200$ keV are subject to significant electron stopping, which does not lead to displaced atoms (see point \# 3 in Sec. \ref{sec:trim}), explaining the spike at $x \approx 200$ nm. The numerical noise at the tail of the neutron distribution arising from the lack of Monte Carlo data should not affect the analysis, as most dark matter signals at these tracks lengths are negligible (absent after binning the distribution) with a reasonable exposure.

We also see a decrease in tracks at low energies due to the $\mathcal{P}_{track}$ suppression across all dark matter masses and backgrounds. At low resolution ($10$ nm), differences at the short-track end fall below the readout threshold; while at high resolution ($1$ nm), they lead to a loss of sensitivity.

\subsubsection{New light mediators}

In the top right, bottom left, and bottom right panels of Figure \ref{fig:dRdx_sp_comparison}, we show the differential track length distribution of solar neutrinos with a nucleus via scalar, vector and axial-vector mediators respectively. In all three scenarios, we compare the standard model (grey) with the following cases: 1) $m_X = 1$ keV (blue), 2) $m_X = 10$ MeV (orange), and 3) $m_X = 1$ GeV (green), where $X=$ scalar $\phi$, vector $Z'$ or axial-vector $a$. These masses are chosen to represent the range of characteristic constraint features, from the insensitivity of cross sections to mediator mass at low masses, to a $\sim 1/m_X^2$ suppression in the cross section for high masses, with corresponding couplings roughly at or an order of magnitude lower than current limits. The blue and green highlighted regions indicate tracks detectable with $10$ nm and $1$ nm resolution readouts respectively. We also show in dashed lines the rate obtained from stopping power-only calculations.

In Sec. \ref{sec:cevens_theory}, we reviewed how the neutrino-nucleus cross section is modified by the addition of new scalar-mediated interactions. We expected such an inclusion would enhance the recoil rate regardless of the properties of the scalar mediator, as given in Eq. \eqref{eq:dsigdE_scalar}. All three masses lead to higher neutrino-nucleus interaction rates than the standard model as expected. Since the cross section scales inversely with $m_\phi^2$, the lower the mass the higher the rate. As shown in the top right panel of Figure \ref{fig:dRdx_sp_comparison}, $m_\phi = 1$ keV consistently has a higher rate than $m_\phi = 10$ MeV at the same coupling. With smaller $m_\phi$, the changes compared to the standard model are more drastic at shorter track lengths than at longer track lengths; whereas for larger $m_\phi$, the increase is more uniform across all track lengths. 

Vector mediators, on the other hand, do not always contribute positively to the differential cross section. The negative interference term in Eq. \eqref{eq:dsigdE_vector} scales inversely with $m_{Z'}^2$ while the positive term scales inversely with $m_{Z'}^4$. Consequently, for small $m_{Z'}$, where both interference terms are non-negligible, the modified rates can be either higher or lower than the standard model depending on the mediator mass, coupling and track length range. This trend is evident in the case of $m_{Z'} = 1$ keV, where the rate exceeds the standard model for $x \lesssim 4$ nm but falls below it for $x \gtrsim 4$ nm. This highlights the importance of scanning the entire track length range, particularly for vector mediators, to capture this distinct profile. For larger $m_{Z'}$, the positive term is suppressed by $m_{Z'}^4$, leading to an overall deficit in the recoil rate with respect to the standard model. For instance, both $m_{Z'} = 10$ MeV and $m_{Z'} = 1$ GeV predict a reduction in solar neutrino recoils.

Unlike the scalar and vector cases, the axial vector contribution has a cubic dependence on the nuclear spin, $S_N^3$, making it an important factor to consider when selecting a target mineral. In olivine, $^{25}$Mg (10\% natural abundance) has spin $S_{N} = 5/2$, $^{29}$Si (4.7\%) and $^{57}$Fe (2.2\%)  both contribute with $S_{N} = 1/2$. Since the positive terms in Eq. \eqref{eq:dsigdE_axialVec} always dominate, the recoil rate is consistently higher than the standard model as shown in Figure \ref{fig:dRdx_sp_comparison}. The overall profile for all three masses closely resembles that of the scalar case. 

\begin{figure}[ht]
    \centering
    \includegraphics[width=\linewidth]{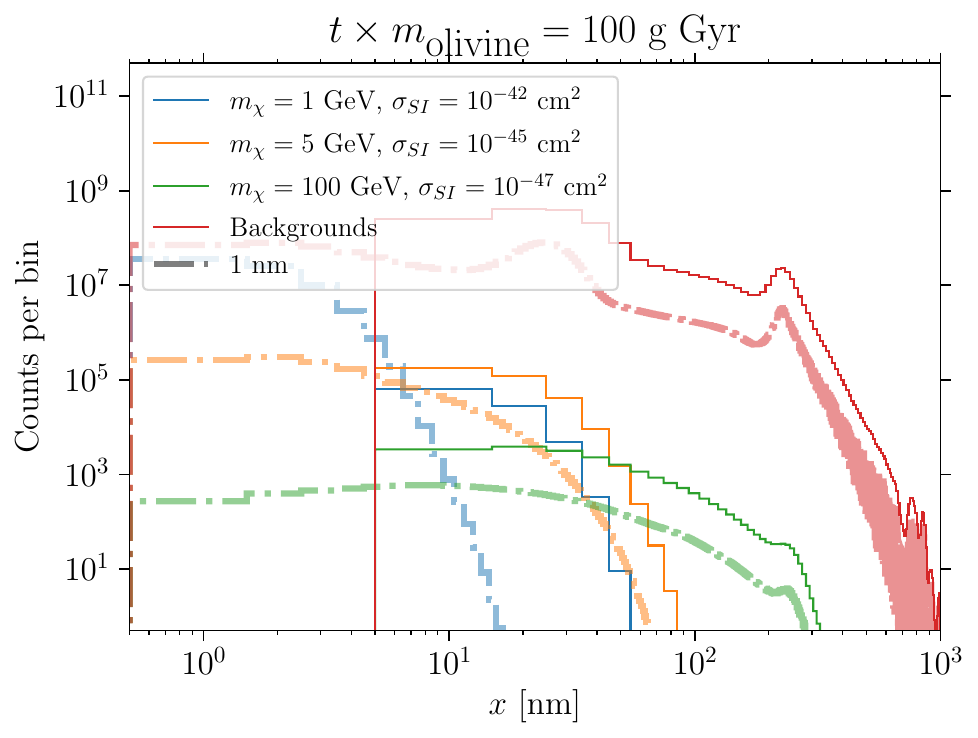}
    \caption{Binned track length distributions for WIMPs with $m_\chi = 1$ GeV, $\sigma_{SI} = 10^{-42}$ cm$^2$; $m_\chi = 5$ GeV, $\sigma_{SI} = 10^{-45}$ cm$^2$; $m_\chi = 100$ GeV, $\sigma_{SI} = 10^{-47}$ cm$^2$, along with background rates from solar neutrinos, atmospheric neutrinos, DSNB neutrinos, GSNB neutrinos and neutrons from U/Th decay. Solid thin lines correspond to a $10$ nm resolution while thick dash-dotted lines correspond to a $1$ nm resolution. Bin widths are chosen to match the resolution, resulting in 100 bins for $10$ nm and 1000 bins for $1$ nm, both with an exposure of $100$ g Gyr.}
    \label{fig:binned_vector_1}
\end{figure}

\section{Sensitivity projections}
\label{sec:limits}
We utilise a binned likelihood fit to assess the sensitivity of paleodetectors. Technologies that could potentially be used for paleodetection, such as scanning electron microscopes (SEM) and transmission electron microscopes (TEM), have readout resolutions of $1 - 10$ nm and sub-nanometer, respectively. While high-resolution readout allows for better identification of signal features, it is time-consuming; lower resolution, on the other hand, could achieve a higher volume of readout in a similar time frame. We therefore consider two cases: one assuming a low-resolution readout with around $100$ g of olivine sample, and the other assuming a high-resolution with around 0.01g.

Due to finite resolution, tracks with true lengths near the edges of each bin may be misclassified into neighbouring bins. To account for this, we follow the approach in \cite{Baum:2021jak}, where the binned distributions are modelled by convolving the differential recoil rate with a window function:
\begin{align}
\label{eq:erf}
W(x'; x_i^{left},& x_i^{right}) = \\
&\frac{1}{2} \left[ \mbox{erf} \left(\frac{x'-x_i^{left}}{\sqrt{2}\sigma_x}\right) - \mbox{erf} \left(\frac{x'-x_i^{right}}{\sqrt{2}\sigma_x}\right) \right]   \nonumber
\end{align}
where $x_i^{left}$ and $x_i^{left}$ are the left and right bin edges respectively, and $\sigma_x$ is the desired resolution, such that 
\begin{equation}
    R_i(x_i^{left}, x_i^{right}) = \int dx' W(x'; x_i^{left}, x_i^{right}) 
    \frac{dR}{dx}(x').
\end{equation}
We obtain limit projections by assuming an Asimov dataset with Standard Model-only signal. Figure \ref{fig:binned_vector_1} shows an example of binned track counts for vector mediator after applying the convolution with Eq. \eqref{eq:erf}. We construct a joint likelihood distribution by treating each bin, as well as selected nuisance parameters, as independent measurements,
\begin{align}
\label{eq:logL_track}
    -2\ln \mathcal{L}(\{\vartheta\} ; \mathbf{\theta_j}) &= \sum_i^{\texttt{N bins}} \frac{(R_{data, i} - R_{th,i}(\{\vartheta\}))^2}{R_{th,i}(\{\vartheta\})} \\
    &+ \sum_j^{\texttt{M}} \frac{(\theta_j - \theta_{central,j})^2}{\sigma_j^2}. \nonumber
\end{align}
The $N$ bins of track counts are summed over, with $R_{data,i}$ representing the counts for $i$-th bin of the Asimov data and $R_{th,i}$ denoting the theoretically expected number of tracks in that bin, evaluated at model parameters $\{\vartheta\}$. In the case of dark matter recoils, for example, $\{\vartheta\} = (m_\chi, \sigma_{SI})$. The index $j$ represents each of the $M$ nuisance parameters, all of which are listed in Table \ref{tab:nuisance}. $\theta_j$ represents the value of the $j$-th nuisance parameter to be tested, with uncertainty $\sigma_j$ and central measured value $\theta_{central,j}$. 

We use a profile likelihood ratio defined as 
\begin{equation}
\label{eq:profile_L}
    \lambda(\{\vartheta\}) = -2 \ln \frac{\mathcal{L}(\vartheta; \hat{\hat{\theta}})}{\mathcal{L}(\hat{\vartheta}; \hat{\mathbf{\theta}})}.
\end{equation}
In the numerator of Eq. \eqref{eq:profile_L}, $\mathcal{L}$ is maximised over the M-dimensional vector $\mathbf{\theta}$ at each point in parameter space, i.e. each pair of ($m_\chi, \sigma_{SI}$) for dark matter scattering, and ($g_{\nu,\phi/Z'},m_{\phi/Z'}$) for light mediators. In the denominator, $\mathcal{L}$ is maximised over the (M+2)-dimensional space $(\vartheta, \mathbf{\theta})$. We exclude parameter values beyond the $90$\% confidence level of the test statistics, such that the pairs of $\vartheta$ values that render $\lambda(\{\vartheta\}) < -2.71$ are rejected. Ref. \cite{Baum:2021jak} explored the sensitivity of projected limits to systematic errors and uranium concentration in more detail. 

\subsection{WIMPs}
\begin{figure*}[!htb]
    \centering
    \includegraphics[width=\textwidth]{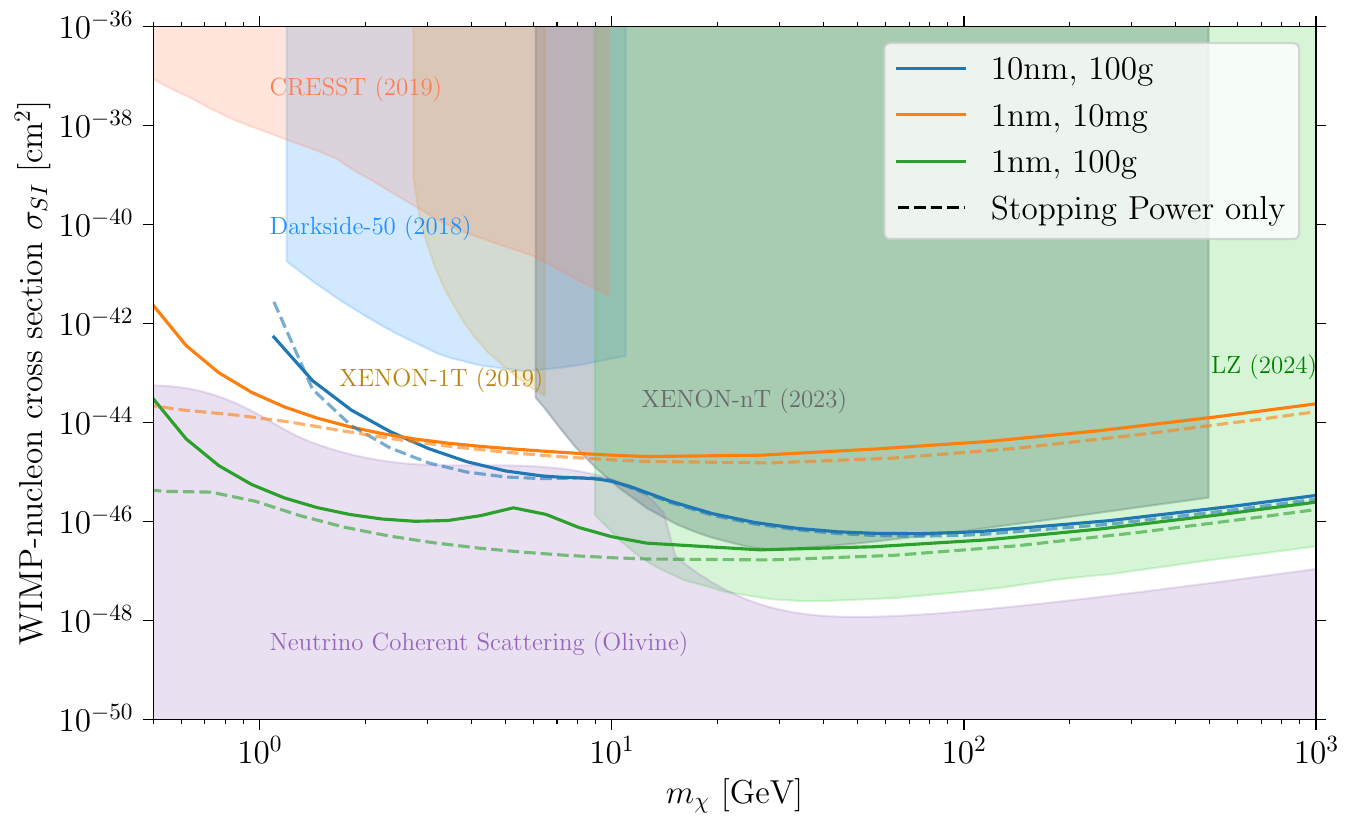}
    \caption{90\% projected sensitivity to spin-independent scattering for WIMP dark matter. All shaded regions represent existing constraints, except for the purple region, which corresponds to the neutrino fog computed for olivine using the NeutrinoFog software \cite{OHare:2021utq}. Three solid lines represent different resolution and target mass scenarios considered, while the dotted lines show corresponding stopping power limits. Other constraints include LZ \cite{LZ:2024zvo}, DarkSide-50 \cite{DarkSide:2018bpj}, CRESST \cite{CRESST:2019jnq}, Xenon-1T \cite{XENON:2019gfn} and XENON-nT \cite{XENON:2023cxc}.}
    \label{fig:limits_wimps}
\end{figure*}

Figure \ref{fig:limits_wimps} shows our projected limits on the WIMP-nucleon cross sections. We present three cases: 1) low readout resolution with a high exposure (LR-HE), i.e., $10$ nm with $100$ g of olivine, 2) high resolution of $1$ nm with a small sample size of $0.01$ g (HR-LE), and 3) high resolution with a large sample size (HR-HE). While scanning $100$ g at $1$ nm would be very ambitious with current technology, we include this case to illustrate the theoretical reach of such paleodetection searches. We also illustrate the neutrino fog for olivine, computed with the help of  the NeutrinoFog\footnote{\url{https://github.com/cajohare/NeutrinoFog}} software \cite{OHare:2021utq}. As explained above, coherent scattering with background neutrinos is accounted for in all of our sensitivity curves, as are neutrons which constitute the bulk of our expected background.

Figure \ref{fig:limits_wimps} compares the limits derived using stopping power (dotted lines) and those obtained from a realistic treatment of track length distribution  (solid lines) across all three readout scenarios. In general, using realistic track distributions yields less stringent limits. This is expected, as our simulations predict a broader track length distribution compared to stopping power, making distinct features of dark matter profile less discernible in the presence of background. At high masses, stopping power is a good approximation since heavy dark matter induces more energetic recoils, where the track length distributions tend to be spiky as shown in Figure \ref{fig:fitting_dist}.   

At low masses however, the discrepancy becomes more pronounced. With HR-LE readout, the sensitivity threshold for low mass WIMPs is approximately half an order of magnitude higher than in the stopping power case. This increase is likely due to the $\mathcal{P}_{track}$ adjustment in track length computation, which suppresses shorter tracks that are primarily expected from light dark matter. It is also worth noting that high resolution improves sensitivity only at low masses, while exposure becomes the limiting factor at high masses. This is consistent with expectations as low mass dark matter primarily produces short tracks while heavy dark matter typically results in tracks well above the threshold of both resolution scenarios. The green lines in Figure \ref{fig:limits_wimps} indicate the sensitivity paleodetection could achieve with HR-HE readout.

Note that the stopping power limits in this work are computed using stopping power tables from \srim rather than directly retrieving them from Ref. \cite{Baum:2021jak}, where the projected upper limits for WIMPs in Olivine were computed. This ensures that the only variable being tested is the inclusion of realistic track distributions, instead of potential discrepancies in assumptions or computational methods. These projected limits assume a 14\% error in solar neutrino fluxes (see Sec. \ref{sec:backgrounds} for details), rather than the 100\% used in Ref. \cite{Baum:2021jak}, and consider a lower resolution of $10$ nm, as opposed to the $15$ nm used in Ref. \cite{Baum:2021jak}.

\begin{figure*}[!htb]
    \includegraphics[width=\textwidth]{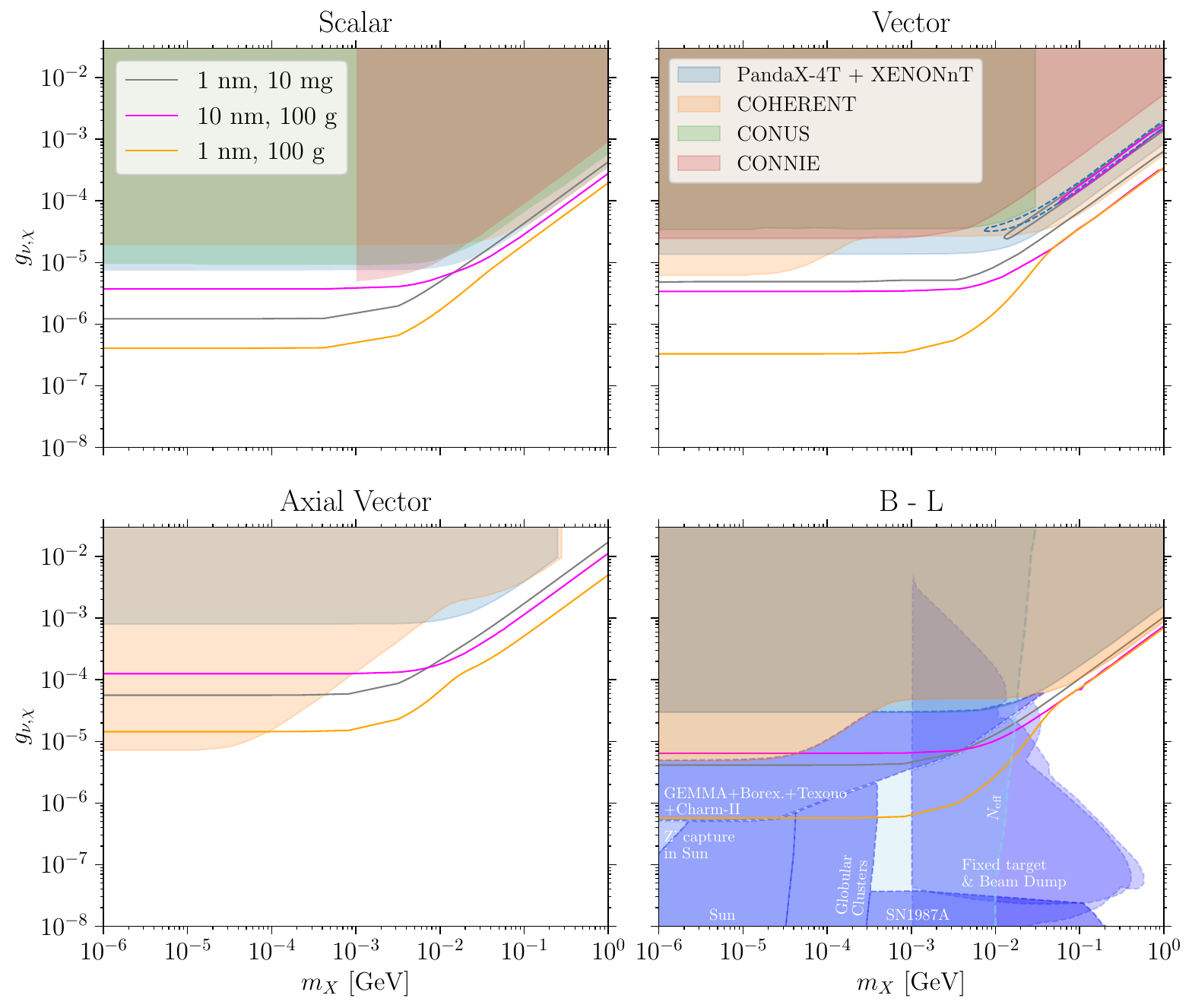}
    \caption{90 \% C.L. projected sensitivities for neutrino-nucleus interactions mediated by light scalar (top left), vector (top right), axial-vector particles (bottom left), and $B-L$ (bottom right). Solid lines represent the three resolution and exposure scenarios considered in this study: magenta corresponds to a threshold of $10$ nm with an exposure of $100$ g Gyr (=$10^5$ ton-year), grey to $1$ nm with 0.01 g Gyr (= $10$ ton-year), and orange to $1$ nm with $100$ g Gyr. The highlighted regions indicated existing limits from COHERENT \cite{DeRomeri:2022twg,AtzoriCorona:2022moj}, CONUS~\cite{CONUS:2021dwh,Lindner:2024eng}, CONNIE~\cite{CONNIE:2019xid, CONNIE:2024pwt}, combined analysis from PanadaX-4T and XENONnT~ \cite{DeRomeri:2024iaw}. Constraints specific to $B-L$ gauge bosons are shown in blue, including combined analaysis from GEMMA, Borexino, TEXONO and Charm II \cite{Bilmis:2015lja}, beam dump and fixed targets experiments (including LSND, MiniBooNE, and NuMI/MINOS \cite{Batell:2009di, Essig:2010gu}, combined from Ref. \cite{DeRomeri:2024iaw}: E141 \cite{Riordan:1987aw}, E137 \cite{Bjorken:1988as}, E774 \cite{Bross:1989mp}, KEK \cite{Konaka:1986cb}, Orsay \cite{Davier:1989wz, Bjorken:2009mm, Andreas:2012mt}, $\nu$-CAL I \cite{Blumlein:1990ay, Blumlein:1991xh, Blumlein:2011mv, Blumlein:2013cua}, CHARM \cite{CHARM:1985anb, Gninenko:2012eq}, NOMAD \cite{NOMAD:2001eyx}, PS191 \cite{Bernardi:1985ny, Gninenko:2011uv}, A1 \cite{Merkel:2014avp}, APEX \cite{APEX:2011dww}), SN1987A supernova neutrinos \cite{Rrapaj:2015wgs}, cooling bounds from globular clusters \cite{Harnik:2012ni}, $N_{\mathrm{eff}}$ bounds from Planck 2018 \cite{Ghosh:2024cxi} and helioscope bounds from the Sun  \cite{Redondo:2008aa}. A small region of the ``Sun" limits could be lifted if a fraction of the emitted bosons are re-absorbed within the Sun \cite{Redondo:2008aa}, this area is indicated by the light blue region near $m_\chi = 10^{-6}$ GeV.}
    \label{fig:limits_light_mediators}
\end{figure*}
\subsection{Solar neutrinos and new light mediators}

In Figure \ref{fig:limits_light_mediators}, we present the projected limits derived from Asimov solar neutrino data in solid lines on the couplings between neutrinos and new scalar, vector and axial-vector mediators as a function of mediator mass. As with the WIMP case, we assess the sensitivity of paleo detectors under three readout scenarios: $10$ nm with $100$ g (LR-HE) in magenta, $1$ nm with $10$ mg (HR-LE) in grey, and $1$ nm with $100$ g (HR-HE) in orange. We also overlaid existing limits, including CE$\nu$NS constraints from COHERENT \cite{DeRomeri:2022twg,AtzoriCorona:2022moj}, CONUS \cite{CONUS:2021dwh,Lindner:2024eng}, CONNIE \cite{CONNIE:2019xid, CONNIE:2024pwt}, and combined analysis using data from PandaX-4T and XENONnT \cite{DeRomeri:2024iaw}.

The shapes of the projected exclusion regions are similar for all three mediators. For $m_X \lesssim 10^{-3}$ GeV, the limits are insensitive to mediator mass, while at higher masses, the sensitivity scales approximately as $m_X^{-2}$ or $m_X^{-4}$ depending on the spin structure of the mediator. 

The LR-HE scenario marginally improves sensitivity beyond existing limits for scalar mediators, while HR-LE extends the reach by an additional order of magnitude. Since light mediators generally produce lower energy recoils, and thus shorter tracks, exposure is less of a limiting factor than readout resolution in contrast to the case with dark matter. A similar trend is observed for axial-vector mediators (bottom left panel of Figure \ref{fig:limits_light_mediators}). Paleodetection with olivine is only half an order of magnitude more sensitive than existing limits. It is however also worth noting that this result depends on the nuclear spins of the target mineral. This means that with an appropriate choice of mineral containing a high abundance of isotopes with large nuclear spins, the sensitivity can be further improved. For vector mediators, the sensitivity is not a monotonic function of mass due to the negative interference contributions in the modified cross section as discussed in Sec. \ref{sec:cevens_theory}. Consequently, a narrow band of allowed region appears within the exclusion bounds. However, this region has already been ruled out by other experiments such as COHERENT.

\subsection{$U(1)_{B-L}$}
\label{sec:limtis_B-L}

The bottom right panel of Figure \ref{fig:limits_light_mediators} shows the $B-L$ parameter space in the gauge coupling $g_{B-L} - m_{Z'}$ plane. The region excluded by both the LR-HE and HR-LE limits has already been heavily constrained down to $g_{\nu, Z'} \approx 10^{-10}$ by neutrino-electron scattering experiments \cite{Bilmis:2015lja}, beam dump and fixed-targets experiments \cite{Bjorken:2009mm,Batell:2009di,Essig:2010gu, DeRomeri:2024iaw, Riordan:1987aw, Bjorken:1988as, Bross:1989mp, Konaka:1986cb, Davier:1989wz, Bjorken:2009mm, Andreas:2012mt, Blumlein:1990ay, Blumlein:1991xh, Blumlein:2011mv, Blumlein:2013cua, CHARM:1985anb, Gninenko:2012eq, NOMAD:2001eyx, Bernardi:1985ny, Gninenko:2011uv, Merkel:2014avp} (see caption of Figure \ref{fig:limits_light_mediators} for detailed references), Supernova 1987A \cite{Rrapaj:2015wgs}, and stellar cooling bounds from globular clusters \cite{Harnik:2012ni} and the Sun as shown in the bottom right panel of Figure \ref{fig:limits_light_mediators}. $B-L$ gauge bosons emitted by the Sun could be reabsorbed by plasmons, reducing the effect of stellar cooling and potentially loosening the constraints \cite{Redondo:2008aa}. This region is labelled as ``$Z'$ capture in Sun". Nonetheless, there is still some potential parameter space to be covered by paleodetection at high masses. The `gap' below $m_{Z'} = 10^{-3}$ GeV is ruled out by the expansion history of the Universe ($N_{\mathrm{eff}}$ as measured by Planck \cite{Planck:2018vyg,Ghosh:2024cxi}), though additional dark radiation in the Early Universe could potentially reopen this parameter space.

\section{Discussion and Conclusion}
\label{sec:conclusion}
We have updated the predicted sensitivities of paleodetectors for the search for particle dark matter, taking into account the realistic track length distributions produced by simulations of particle propagation in medium. We have also re-examined the spectrum expected from solar neutrinos, and provided sensitivities to new light mediators which may enhance or suppress the track production rate at low recoil energies.

While the $\sim$ g Gyr exposures here are not unrealistic, the sheer number of track measurements that they imply should give the reader pause. The O($10^7$) tracks per bin shown in e.g. Figure \ref{fig:dRdx_sp_comparison} from background alone is consistent with predictions from previous work. Preliminary measurements in halite at the University of North Florida (Sec 3 of \cite{Baum:2024eyr}) recorded a few hundred tracks using Ar plasma etching and confocal microscopy, consistent with the Th-$\alpha$ peak plus neutron background predicted in Ref. \cite{Drukier:2018pdy} (the same neutron model used here). These measurements imply surface track densities of $\sim 10^7$ cm$^{-2}$. 

Predictions thus appear somewhat robust, implying that the development of technology to read out such large track numbers is imperative. Many efforts are underway, using automated search algorithms and technology such as atomic force microscopy, etching, and X-ray diffraction \cite{Baum:2024eyr}, but scalability remains an open question. 

Nonetheless, as traditional direct detection experiments pass the 10-ton scale and begin to require significant funding and stockpiling of target material, paleodetection will likely become the fastest---and cheapest---path beyond the 100 ton-year mark. As we have shown here, the path through the neutrino fog may indeed come from deeper underground than Gran Sasso, SNOLAB, or SURF.

\begin{acknowledgments}
We thank Joe Bramante, Yilda Boukhtouchen and Sharlotte Mkhonto from the QCUMBER collaboration, as well as Neal Avis Kozar, Chris Cappiello and Avi Friedlander for helpful input. ACV is grateful to members of the LPPC (Harvard) and CTP (MIT) for their hospitality while completing this work. This work was supported by Arthur B. McDonald Canadian Astroparticle Physics Research Institute, NSERC the Canada Foundation for Innovation and the Province of Ontario. Research at Perimeter Institute is supported by the Government of Canada through the Department of Innovation, Science, and Economic Development, and by the Province of Ontario. 
\end{acknowledgments}

\FloatBarrier

\bibliographystyle{apsrev4-1}
\bibliography{Bibli}

\begin{thebibliography}{85}%
\makeatletter
\providecommand \@ifxundefined [1]{%
 \@ifx{#1\undefined}
}%
\providecommand \@ifnum [1]{%
 \ifnum #1\expandafter \@firstoftwo
 \else \expandafter \@secondoftwo
 \fi
}%
\providecommand \@ifx [1]{%
 \ifx #1\expandafter \@firstoftwo
 \else \expandafter \@secondoftwo
 \fi
}%
\providecommand \natexlab [1]{#1}%
\providecommand \enquote  [1]{``#1''}%
\providecommand \bibnamefont  [1]{#1}%
\providecommand \bibfnamefont [1]{#1}%
\providecommand \citenamefont [1]{#1}%
\providecommand \href@noop [0]{\@secondoftwo}%
\providecommand \href [0]{\begingroup \@sanitize@url \@href}%
\providecommand \@href[1]{\@@startlink{#1}\@@href}%
\providecommand \@@href[1]{\endgroup#1\@@endlink}%
\providecommand \@sanitize@url [0]{\catcode `\\12\catcode `\$12\catcode
  `\&12\catcode `\#12\catcode `\^12\catcode `\_12\catcode `\%12\relax}%
\providecommand \@@startlink[1]{}%
\providecommand \@@endlink[0]{}%
\providecommand \url  [0]{\begingroup\@sanitize@url \@url }%
\providecommand \@url [1]{\endgroup\@href {#1}{\urlprefix }}%
\providecommand \urlprefix  [0]{URL }%
\providecommand \Eprint [0]{\href }%
\providecommand \doibase [0]{http://dx.doi.org/}%
\providecommand \selectlanguage [0]{\@gobble}%
\providecommand \bibinfo  [0]{\@secondoftwo}%
\providecommand \bibfield  [0]{\@secondoftwo}%
\providecommand \translation [1]{[#1]}%
\providecommand \BibitemOpen [0]{}%
\providecommand \bibitemStop [0]{}%
\providecommand \bibitemNoStop [0]{.\EOS\space}%
\providecommand \EOS [0]{\spacefactor3000\relax}%
\providecommand \BibitemShut  [1]{\csname bibitem#1\endcsname}%
\let\auto@bib@innerbib\@empty
\bibitem [{\citenamefont {Aghanim}\ \emph {et~al.}(2020)\citenamefont {Aghanim}
  \emph {et~al.}}]{Planck:2018vyg}%
  \BibitemOpen
  \bibfield  {author} {\bibinfo {author} {\bibfnamefont {N.}~\bibnamefont
  {Aghanim}} \emph {et~al.} (\bibinfo {collaboration} {Planck}),\ }\href
  {\doibase 10.1051/0004-6361/201833910} {\bibfield  {journal} {\bibinfo
  {journal} {Astron. Astrophys.}\ }\textbf {\bibinfo {volume} {641}},\ \bibinfo
  {pages} {A6} (\bibinfo {year} {2020})},\ \bibinfo {note} {[Erratum:
  Astron.Astrophys. 652, C4 (2021)]},\ \Eprint
  {http://arxiv.org/abs/1807.06209} {arXiv:1807.06209 [astro-ph.CO]}
  \BibitemShut {NoStop}%
\bibitem [{\citenamefont {Aalbers}\ \emph
  {et~al.}(2024{\natexlab{a}})\citenamefont {Aalbers} \emph
  {et~al.}}]{LZCollaboration:2024lux}%
  \BibitemOpen
  \bibfield  {author} {\bibinfo {author} {\bibfnamefont {J.}~\bibnamefont
  {Aalbers}} \emph {et~al.} (\bibinfo {collaboration} {LZ Collaboration}),\
  }\href@noop {} {\  (\bibinfo {year} {2024}{\natexlab{a}})},\ \Eprint
  {http://arxiv.org/abs/2410.17036} {arXiv:2410.17036 [hep-ex]} \BibitemShut
  {NoStop}%
\bibitem [{\citenamefont {Amole}\ \emph {et~al.}(2019)\citenamefont {Amole}
  \emph {et~al.}}]{PICO:2019vsc}%
  \BibitemOpen
  \bibfield  {author} {\bibinfo {author} {\bibfnamefont {C.}~\bibnamefont
  {Amole}} \emph {et~al.} (\bibinfo {collaboration} {PICO}),\ }\href {\doibase
  10.1103/PhysRevD.100.022001} {\bibfield  {journal} {\bibinfo  {journal}
  {Phys. Rev. D}\ }\textbf {\bibinfo {volume} {100}},\ \bibinfo {pages}
  {022001} (\bibinfo {year} {2019})},\ \Eprint
  {http://arxiv.org/abs/1902.04031} {arXiv:1902.04031 [astro-ph.CO]}
  \BibitemShut {NoStop}%
\bibitem [{\citenamefont {Agnes}\ \emph {et~al.}(2023)\citenamefont {Agnes}
  \emph {et~al.}}]{DarkSide-50:2022qzh}%
  \BibitemOpen
  \bibfield  {author} {\bibinfo {author} {\bibfnamefont {P.}~\bibnamefont
  {Agnes}} \emph {et~al.} (\bibinfo {collaboration} {DarkSide-50}),\ }\href
  {\doibase 10.1103/PhysRevD.107.063001} {\bibfield  {journal} {\bibinfo
  {journal} {Phys. Rev. D}\ }\textbf {\bibinfo {volume} {107}},\ \bibinfo
  {pages} {063001} (\bibinfo {year} {2023})},\ \Eprint
  {http://arxiv.org/abs/2207.11966} {arXiv:2207.11966 [hep-ex]} \BibitemShut
  {NoStop}%
\bibitem [{\citenamefont {Arora}\ \emph {et~al.}(2024)\citenamefont {Arora}
  \emph {et~al.}}]{NEWS-G:2024jms}%
  \BibitemOpen
  \bibfield  {author} {\bibinfo {author} {\bibfnamefont {M.~M.}\ \bibnamefont
  {Arora}} \emph {et~al.} (\bibinfo {collaboration} {NEWS-G}),\ }\href@noop {}
  {\  (\bibinfo {year} {2024})},\ \Eprint {http://arxiv.org/abs/2407.12769}
  {arXiv:2407.12769 [hep-ex]} \BibitemShut {NoStop}%
\bibitem [{\citenamefont {Aprile}\ \emph {et~al.}(2023)\citenamefont {Aprile}
  \emph {et~al.}}]{XENON:2023cxc}%
  \BibitemOpen
  \bibfield  {author} {\bibinfo {author} {\bibfnamefont {E.}~\bibnamefont
  {Aprile}} \emph {et~al.} (\bibinfo {collaboration} {XENON}),\ }\href
  {\doibase 10.1103/PhysRevLett.131.041003} {\bibfield  {journal} {\bibinfo
  {journal} {Phys. Rev. Lett.}\ }\textbf {\bibinfo {volume} {131}},\ \bibinfo
  {pages} {041003} (\bibinfo {year} {2023})},\ \Eprint
  {http://arxiv.org/abs/2303.14729} {arXiv:2303.14729 [hep-ex]} \BibitemShut
  {NoStop}%
\bibitem [{\citenamefont {Price}\ and\ \citenamefont
  {Salamon}(1986)}]{Price:1986ky}%
  \BibitemOpen
  \bibfield  {author} {\bibinfo {author} {\bibfnamefont {P.~B.}\ \bibnamefont
  {Price}}\ and\ \bibinfo {author} {\bibfnamefont {M.~H.}\ \bibnamefont
  {Salamon}},\ }\href {\doibase 10.1103/PhysRevLett.56.1226} {\bibfield
  {journal} {\bibinfo  {journal} {Phys. Rev. Lett.}\ }\textbf {\bibinfo
  {volume} {56}},\ \bibinfo {pages} {1226} (\bibinfo {year}
  {1986})}\BibitemShut {NoStop}%
\bibitem [{\citenamefont {Snowden-Ifft}\ \emph {et~al.}(1995)\citenamefont
  {Snowden-Ifft}, \citenamefont {Freeman},\ and\ \citenamefont
  {Price}}]{Snowden-Ifft:1995zgn}%
  \BibitemOpen
  \bibfield  {author} {\bibinfo {author} {\bibfnamefont {D.~P.}\ \bibnamefont
  {Snowden-Ifft}}, \bibinfo {author} {\bibfnamefont {E.~S.}\ \bibnamefont
  {Freeman}}, \ and\ \bibinfo {author} {\bibfnamefont {P.~B.}\ \bibnamefont
  {Price}},\ }\href {\doibase 10.1103/PhysRevLett.74.4133} {\bibfield
  {journal} {\bibinfo  {journal} {Phys. Rev. Lett.}\ }\textbf {\bibinfo
  {volume} {74}},\ \bibinfo {pages} {4133} (\bibinfo {year}
  {1995})}\BibitemShut {NoStop}%
\bibitem [{\citenamefont {Acevedo}\ \emph {et~al.}(2023)\citenamefont
  {Acevedo}, \citenamefont {Bramante},\ and\ \citenamefont
  {Goodman}}]{Acevedo:2021tbl}%
  \BibitemOpen
  \bibfield  {author} {\bibinfo {author} {\bibfnamefont {J.~F.}\ \bibnamefont
  {Acevedo}}, \bibinfo {author} {\bibfnamefont {J.}~\bibnamefont {Bramante}}, \
  and\ \bibinfo {author} {\bibfnamefont {A.}~\bibnamefont {Goodman}},\ }\href
  {\doibase 10.1088/1475-7516/2023/11/085} {\bibfield  {journal} {\bibinfo
  {journal} {JCAP}\ }\textbf {\bibinfo {volume} {11}},\ \bibinfo {pages} {085}
  (\bibinfo {year} {2023})},\ \Eprint {http://arxiv.org/abs/2105.06473}
  {arXiv:2105.06473 [hep-ph]} \BibitemShut {NoStop}%
\bibitem [{\citenamefont {Baum}\ \emph
  {et~al.}(2020{\natexlab{a}})\citenamefont {Baum}, \citenamefont {Drukier},
  \citenamefont {Freese}, \citenamefont {G\'orski},\ and\ \citenamefont
  {Stengel}}]{Baum:2018tfw}%
  \BibitemOpen
  \bibfield  {author} {\bibinfo {author} {\bibfnamefont {S.}~\bibnamefont
  {Baum}}, \bibinfo {author} {\bibfnamefont {A.~K.}\ \bibnamefont {Drukier}},
  \bibinfo {author} {\bibfnamefont {K.}~\bibnamefont {Freese}}, \bibinfo
  {author} {\bibfnamefont {M.}~\bibnamefont {G\'orski}}, \ and\ \bibinfo
  {author} {\bibfnamefont {P.}~\bibnamefont {Stengel}},\ }\href {\doibase
  10.1016/j.physletb.2020.135325} {\bibfield  {journal} {\bibinfo  {journal}
  {Phys. Lett. B}\ }\textbf {\bibinfo {volume} {803}},\ \bibinfo {pages}
  {135325} (\bibinfo {year} {2020}{\natexlab{a}})},\ \Eprint
  {http://arxiv.org/abs/1806.05991} {arXiv:1806.05991 [astro-ph.CO]}
  \BibitemShut {NoStop}%
\bibitem [{\citenamefont {Edwards}\ \emph {et~al.}(2019)\citenamefont
  {Edwards}, \citenamefont {Kavanagh}, \citenamefont {Weniger}, \citenamefont
  {Baum}, \citenamefont {Drukier}, \citenamefont {Freese}, \citenamefont
  {G\'orski},\ and\ \citenamefont {Stengel}}]{Edwards:2018hcf}%
  \BibitemOpen
  \bibfield  {author} {\bibinfo {author} {\bibfnamefont {T.~D.~P.}\
  \bibnamefont {Edwards}}, \bibinfo {author} {\bibfnamefont {B.~J.}\
  \bibnamefont {Kavanagh}}, \bibinfo {author} {\bibfnamefont {C.}~\bibnamefont
  {Weniger}}, \bibinfo {author} {\bibfnamefont {S.}~\bibnamefont {Baum}},
  \bibinfo {author} {\bibfnamefont {A.~K.}\ \bibnamefont {Drukier}}, \bibinfo
  {author} {\bibfnamefont {K.}~\bibnamefont {Freese}}, \bibinfo {author}
  {\bibfnamefont {M.}~\bibnamefont {G\'orski}}, \ and\ \bibinfo {author}
  {\bibfnamefont {P.}~\bibnamefont {Stengel}},\ }\href {\doibase
  10.1103/PhysRevD.99.043541} {\bibfield  {journal} {\bibinfo  {journal} {Phys.
  Rev. D}\ }\textbf {\bibinfo {volume} {99}},\ \bibinfo {pages} {043541}
  (\bibinfo {year} {2019})},\ \Eprint {http://arxiv.org/abs/1811.10549}
  {arXiv:1811.10549 [hep-ph]} \BibitemShut {NoStop}%
\bibitem [{\citenamefont {Drukier}\ \emph {et~al.}(2019)\citenamefont
  {Drukier}, \citenamefont {Baum}, \citenamefont {Freese}, \citenamefont
  {G\'orski},\ and\ \citenamefont {Stengel}}]{Drukier:2018pdy}%
  \BibitemOpen
  \bibfield  {author} {\bibinfo {author} {\bibfnamefont {A.~K.}\ \bibnamefont
  {Drukier}}, \bibinfo {author} {\bibfnamefont {S.}~\bibnamefont {Baum}},
  \bibinfo {author} {\bibfnamefont {K.}~\bibnamefont {Freese}}, \bibinfo
  {author} {\bibfnamefont {M.}~\bibnamefont {G\'orski}}, \ and\ \bibinfo
  {author} {\bibfnamefont {P.}~\bibnamefont {Stengel}},\ }\href {\doibase
  10.1103/PhysRevD.99.043014} {\bibfield  {journal} {\bibinfo  {journal} {Phys.
  Rev. D}\ }\textbf {\bibinfo {volume} {99}},\ \bibinfo {pages} {043014}
  (\bibinfo {year} {2019})},\ \Eprint {http://arxiv.org/abs/1811.06844}
  {arXiv:1811.06844 [astro-ph.CO]} \BibitemShut {NoStop}%
\bibitem [{\citenamefont {Baum}\ \emph
  {et~al.}(2021{\natexlab{a}})\citenamefont {Baum}, \citenamefont {DeRocco},
  \citenamefont {Edwards},\ and\ \citenamefont {Kalia}}]{Baum:2021chx}%
  \BibitemOpen
  \bibfield  {author} {\bibinfo {author} {\bibfnamefont {S.}~\bibnamefont
  {Baum}}, \bibinfo {author} {\bibfnamefont {W.}~\bibnamefont {DeRocco}},
  \bibinfo {author} {\bibfnamefont {T.~D.~P.}\ \bibnamefont {Edwards}}, \ and\
  \bibinfo {author} {\bibfnamefont {S.}~\bibnamefont {Kalia}},\ }\href
  {\doibase 10.1103/PhysRevD.104.123015} {\bibfield  {journal} {\bibinfo
  {journal} {Phys. Rev. D}\ }\textbf {\bibinfo {volume} {104}},\ \bibinfo
  {pages} {123015} (\bibinfo {year} {2021}{\natexlab{a}})},\ \Eprint
  {http://arxiv.org/abs/2107.02812} {arXiv:2107.02812 [astro-ph.GA]}
  \BibitemShut {NoStop}%
\bibitem [{\citenamefont {Baum}\ \emph
  {et~al.}(2021{\natexlab{b}})\citenamefont {Baum}, \citenamefont {Edwards},
  \citenamefont {Freese},\ and\ \citenamefont {Stengel}}]{Baum:2021jak}%
  \BibitemOpen
  \bibfield  {author} {\bibinfo {author} {\bibfnamefont {S.}~\bibnamefont
  {Baum}}, \bibinfo {author} {\bibfnamefont {T.~D.~P.}\ \bibnamefont
  {Edwards}}, \bibinfo {author} {\bibfnamefont {K.}~\bibnamefont {Freese}}, \
  and\ \bibinfo {author} {\bibfnamefont {P.}~\bibnamefont {Stengel}},\ }\href
  {\doibase 10.3390/instruments5020021} {\bibfield  {journal} {\bibinfo
  {journal} {Instruments}\ }\textbf {\bibinfo {volume} {5}},\ \bibinfo {pages}
  {21} (\bibinfo {year} {2021}{\natexlab{b}})},\ \Eprint
  {http://arxiv.org/abs/2106.06559} {arXiv:2106.06559 [astro-ph.CO]}
  \BibitemShut {NoStop}%
\bibitem [{\citenamefont {Baum}\ \emph {et~al.}(2024)\citenamefont {Baum} \emph
  {et~al.}}]{Baum:2024eyr}%
  \BibitemOpen
  \bibfield  {author} {\bibinfo {author} {\bibfnamefont {S.}~\bibnamefont
  {Baum}} \emph {et~al.}\ }(\bibinfo {year} {2024})\ \Eprint
  {http://arxiv.org/abs/2405.01626} {arXiv:2405.01626 [astro-ph.CO]}
  \BibitemShut {NoStop}%
\bibitem [{\citenamefont {Baum}\ \emph
  {et~al.}(2020{\natexlab{b}})\citenamefont {Baum}, \citenamefont {Edwards},
  \citenamefont {Kavanagh}, \citenamefont {Stengel}, \citenamefont {Drukier},
  \citenamefont {Freese}, \citenamefont {G\'orski},\ and\ \citenamefont
  {Weniger}}]{Baum:2019fqm}%
  \BibitemOpen
  \bibfield  {author} {\bibinfo {author} {\bibfnamefont {S.}~\bibnamefont
  {Baum}}, \bibinfo {author} {\bibfnamefont {T.~D.~P.}\ \bibnamefont
  {Edwards}}, \bibinfo {author} {\bibfnamefont {B.~J.}\ \bibnamefont
  {Kavanagh}}, \bibinfo {author} {\bibfnamefont {P.}~\bibnamefont {Stengel}},
  \bibinfo {author} {\bibfnamefont {A.~K.}\ \bibnamefont {Drukier}}, \bibinfo
  {author} {\bibfnamefont {K.}~\bibnamefont {Freese}}, \bibinfo {author}
  {\bibfnamefont {M.}~\bibnamefont {G\'orski}}, \ and\ \bibinfo {author}
  {\bibfnamefont {C.}~\bibnamefont {Weniger}},\ }\href {\doibase
  10.1103/PhysRevD.101.103017} {\bibfield  {journal} {\bibinfo  {journal}
  {Phys. Rev. D}\ }\textbf {\bibinfo {volume} {101}},\ \bibinfo {pages}
  {103017} (\bibinfo {year} {2020}{\natexlab{b}})},\ \Eprint
  {http://arxiv.org/abs/1906.05800} {arXiv:1906.05800 [astro-ph.GA]}
  \BibitemShut {NoStop}%
\bibitem [{\citenamefont {Baum}\ \emph {et~al.}(2022)\citenamefont {Baum},
  \citenamefont {Capozzi},\ and\ \citenamefont {Horiuchi}}]{Baum:2022wfc}%
  \BibitemOpen
  \bibfield  {author} {\bibinfo {author} {\bibfnamefont {S.}~\bibnamefont
  {Baum}}, \bibinfo {author} {\bibfnamefont {F.}~\bibnamefont {Capozzi}}, \
  and\ \bibinfo {author} {\bibfnamefont {S.}~\bibnamefont {Horiuchi}},\ }\href
  {\doibase 10.1103/PhysRevD.106.123008} {\bibfield  {journal} {\bibinfo
  {journal} {Phys. Rev. D}\ }\textbf {\bibinfo {volume} {106}},\ \bibinfo
  {pages} {123008} (\bibinfo {year} {2022})},\ \Eprint
  {http://arxiv.org/abs/2203.12696} {arXiv:2203.12696 [hep-ph]} \BibitemShut
  {NoStop}%
\bibitem [{\citenamefont {Galelli}\ \emph {et~al.}(2023)\citenamefont
  {Galelli}, \citenamefont {Caccianiga}, \citenamefont {Veutro},\ and\
  \citenamefont {Apollonio}}]{Galelli:2023tay}%
  \BibitemOpen
  \bibfield  {author} {\bibinfo {author} {\bibfnamefont {C.}~\bibnamefont
  {Galelli}}, \bibinfo {author} {\bibfnamefont {L.}~\bibnamefont {Caccianiga}},
  \bibinfo {author} {\bibfnamefont {A.}~\bibnamefont {Veutro}}, \ and\ \bibinfo
  {author} {\bibfnamefont {L.}~\bibnamefont {Apollonio}},\ }\href {\doibase
  10.22323/1.423.0145} {\bibfield  {journal} {\bibinfo  {journal} {PoS}\
  }\textbf {\bibinfo {volume} {ECRS}},\ \bibinfo {pages} {145} (\bibinfo {year}
  {2023})}\BibitemShut {NoStop}%
\bibitem [{\citenamefont {Caccianiga}\ \emph {et~al.}(2024)\citenamefont
  {Caccianiga}, \citenamefont {Apollonio}, \citenamefont {Mariani},
  \citenamefont {Magnani}, \citenamefont {Galelli},\ and\ \citenamefont
  {Veutro}}]{Caccianiga:2024otm}%
  \BibitemOpen
  \bibfield  {author} {\bibinfo {author} {\bibfnamefont {L.}~\bibnamefont
  {Caccianiga}}, \bibinfo {author} {\bibfnamefont {L.}~\bibnamefont
  {Apollonio}}, \bibinfo {author} {\bibfnamefont {F.~M.}\ \bibnamefont
  {Mariani}}, \bibinfo {author} {\bibfnamefont {P.}~\bibnamefont {Magnani}},
  \bibinfo {author} {\bibfnamefont {C.}~\bibnamefont {Galelli}}, \ and\
  \bibinfo {author} {\bibfnamefont {A.}~\bibnamefont {Veutro}},\ }\href
  {\doibase 10.1103/PhysRevD.110.L121301} {\bibfield  {journal} {\bibinfo
  {journal} {Phys. Rev. D}\ }\textbf {\bibinfo {volume} {110}},\ \bibinfo
  {pages} {L121301} (\bibinfo {year} {2024})},\ \Eprint
  {http://arxiv.org/abs/2405.04908} {arXiv:2405.04908 [astro-ph.HE]}
  \BibitemShut {NoStop}%
\bibitem [{\citenamefont {Mariani}\ \emph {et~al.}(2023)\citenamefont
  {Mariani}, \citenamefont {Caccianiga}, \citenamefont {Galelli}, \citenamefont
  {Apollonio}, \citenamefont {Magnani},\ and\ \citenamefont
  {Veutro}}]{Mariani:2023mwb}%
  \BibitemOpen
  \bibfield  {author} {\bibinfo {author} {\bibfnamefont {F.~M.}\ \bibnamefont
  {Mariani}}, \bibinfo {author} {\bibfnamefont {L.}~\bibnamefont {Caccianiga}},
  \bibinfo {author} {\bibfnamefont {C.}~\bibnamefont {Galelli}}, \bibinfo
  {author} {\bibfnamefont {L.}~\bibnamefont {Apollonio}}, \bibinfo {author}
  {\bibfnamefont {P.}~\bibnamefont {Magnani}}, \ and\ \bibinfo {author}
  {\bibfnamefont {A.}~\bibnamefont {Veutro}},\ }\href {\doibase
  10.22323/1.444.0544} {\bibfield  {journal} {\bibinfo  {journal} {PoS}\
  }\textbf {\bibinfo {volume} {ICRC2023}},\ \bibinfo {pages} {544} (\bibinfo
  {year} {2023})}\BibitemShut {NoStop}%
\bibitem [{\citenamefont {Jordan}\ \emph {et~al.}(2020)\citenamefont {Jordan},
  \citenamefont {Baum}, \citenamefont {Stengel}, \citenamefont {Ferrari},
  \citenamefont {Morone}, \citenamefont {Sala},\ and\ \citenamefont
  {Spitz}}]{Jordan:2020gxx}%
  \BibitemOpen
  \bibfield  {author} {\bibinfo {author} {\bibfnamefont {J.~R.}\ \bibnamefont
  {Jordan}}, \bibinfo {author} {\bibfnamefont {S.}~\bibnamefont {Baum}},
  \bibinfo {author} {\bibfnamefont {P.}~\bibnamefont {Stengel}}, \bibinfo
  {author} {\bibfnamefont {A.}~\bibnamefont {Ferrari}}, \bibinfo {author}
  {\bibfnamefont {M.~C.}\ \bibnamefont {Morone}}, \bibinfo {author}
  {\bibfnamefont {P.}~\bibnamefont {Sala}}, \ and\ \bibinfo {author}
  {\bibfnamefont {J.}~\bibnamefont {Spitz}},\ }\href {\doibase
  10.1103/PhysRevLett.125.231802} {\bibfield  {journal} {\bibinfo  {journal}
  {Phys. Rev. Lett.}\ }\textbf {\bibinfo {volume} {125}},\ \bibinfo {pages}
  {231802} (\bibinfo {year} {2020})},\ \Eprint
  {http://arxiv.org/abs/2004.08394} {arXiv:2004.08394 [hep-ph]} \BibitemShut
  {NoStop}%
\bibitem [{\citenamefont {Tapia-Arellano}\ and\ \citenamefont
  {Horiuchi}(2021)}]{Tapia-Arellano:2021cml}%
  \BibitemOpen
  \bibfield  {author} {\bibinfo {author} {\bibfnamefont {N.}~\bibnamefont
  {Tapia-Arellano}}\ and\ \bibinfo {author} {\bibfnamefont {S.}~\bibnamefont
  {Horiuchi}},\ }\href {\doibase 10.1103/PhysRevD.103.123016} {\bibfield
  {journal} {\bibinfo  {journal} {Phys. Rev. D}\ }\textbf {\bibinfo {volume}
  {103}},\ \bibinfo {pages} {123016} (\bibinfo {year} {2021})},\ \Eprint
  {http://arxiv.org/abs/2102.01755} {arXiv:2102.01755 [hep-ph]} \BibitemShut
  {NoStop}%
\bibitem [{\citenamefont {Baum}\ \emph {et~al.}(2023)\citenamefont {Baum} \emph
  {et~al.}}]{Baum:2023cct}%
  \BibitemOpen
  \bibfield  {author} {\bibinfo {author} {\bibfnamefont {S.}~\bibnamefont
  {Baum}} \emph {et~al.},\ }\href {\doibase 10.1016/j.dark.2023.101245}
  {\bibfield  {journal} {\bibinfo  {journal} {Phys. Dark Univ.}\ }\textbf
  {\bibinfo {volume} {41}},\ \bibinfo {pages} {101245} (\bibinfo {year}
  {2023})},\ \Eprint {http://arxiv.org/abs/2301.07118} {arXiv:2301.07118
  [astro-ph.IM]} \BibitemShut {NoStop}%
\bibitem [{\citenamefont {Collar}(1996)}]{Collar:1995aw}%
  \BibitemOpen
  \bibfield  {author} {\bibinfo {author} {\bibfnamefont {J.~I.}\ \bibnamefont
  {Collar}},\ }\href {\doibase 10.1103/PhysRevLett.76.331} {\bibfield
  {journal} {\bibinfo  {journal} {Phys. Rev. Lett.}\ }\textbf {\bibinfo
  {volume} {76}},\ \bibinfo {pages} {331} (\bibinfo {year} {1996})},\ \Eprint
  {http://arxiv.org/abs/astro-ph/9511055} {arXiv:astro-ph/9511055} \BibitemShut
  {NoStop}%
\bibitem [{\citenamefont {Ziegler}\ \emph {et~al.}(2010)\citenamefont
  {Ziegler}, \citenamefont {Ziegler},\ and\ \citenamefont
  {Biersack}}]{Ziegler:2010bzy}%
  \BibitemOpen
  \bibfield  {author} {\bibinfo {author} {\bibfnamefont {J.~F.}\ \bibnamefont
  {Ziegler}}, \bibinfo {author} {\bibfnamefont {M.~D.}\ \bibnamefont
  {Ziegler}}, \ and\ \bibinfo {author} {\bibfnamefont {J.~P.}\ \bibnamefont
  {Biersack}},\ }\href {\doibase 10.1016/j.nimb.2010.02.091} {\bibfield
  {journal} {\bibinfo  {journal} {Nucl. Instrum. Meth. B}\ }\textbf {\bibinfo
  {volume} {268}},\ \bibinfo {pages} {1818} (\bibinfo {year}
  {2010})}\BibitemShut {NoStop}%
\bibitem [{\citenamefont {Cerdeno}\ \emph {et~al.}(2016)\citenamefont
  {Cerdeno}, \citenamefont {Fairbairn}, \citenamefont {Jubb}, \citenamefont
  {Machado}, \citenamefont {Vincent},\ and\ \citenamefont
  {B{\oe}hm}}]{cerdeno2016physics}%
  \BibitemOpen
  \bibfield  {author} {\bibinfo {author} {\bibfnamefont {D.~G.}\ \bibnamefont
  {Cerdeno}}, \bibinfo {author} {\bibfnamefont {M.}~\bibnamefont {Fairbairn}},
  \bibinfo {author} {\bibfnamefont {T.}~\bibnamefont {Jubb}}, \bibinfo {author}
  {\bibfnamefont {P.~A.}\ \bibnamefont {Machado}}, \bibinfo {author}
  {\bibfnamefont {A.~C.}\ \bibnamefont {Vincent}}, \ and\ \bibinfo {author}
  {\bibfnamefont {C.}~\bibnamefont {B{\oe}hm}},\ }\href@noop {} {\bibfield
  {journal} {\bibinfo  {journal} {Journal of High Energy Physics}\ }\textbf
  {\bibinfo {volume} {2016}},\ \bibinfo {pages} {1} (\bibinfo {year}
  {2016})}\BibitemShut {NoStop}%
\bibitem [{\citenamefont {De~Romeri}\ \emph
  {et~al.}(2024{\natexlab{a}})\citenamefont {De~Romeri}, \citenamefont
  {Papoulias},\ and\ \citenamefont {Ternes}}]{DeRomeri:2024dbv}%
  \BibitemOpen
  \bibfield  {author} {\bibinfo {author} {\bibfnamefont {V.}~\bibnamefont
  {De~Romeri}}, \bibinfo {author} {\bibfnamefont {D.~K.}\ \bibnamefont
  {Papoulias}}, \ and\ \bibinfo {author} {\bibfnamefont {C.~A.}\ \bibnamefont
  {Ternes}},\ }\href {\doibase 10.1007/JHEP05(2024)165} {\bibfield  {journal}
  {\bibinfo  {journal} {JHEP}\ }\textbf {\bibinfo {volume} {05}},\ \bibinfo
  {pages} {165} (\bibinfo {year} {2024}{\natexlab{a}})},\ \Eprint
  {http://arxiv.org/abs/2402.05506} {arXiv:2402.05506 [hep-ph]} \BibitemShut
  {NoStop}%
\bibitem [{\citenamefont {Evans}\ \emph {et~al.}(2019)\citenamefont {Evans},
  \citenamefont {O'Hare},\ and\ \citenamefont {McCabe}}]{Evans:2019PRD}%
  \BibitemOpen
  \bibfield  {author} {\bibinfo {author} {\bibfnamefont {N.~W.}\ \bibnamefont
  {Evans}}, \bibinfo {author} {\bibfnamefont {C.~A.~J.}\ \bibnamefont
  {O'Hare}}, \ and\ \bibinfo {author} {\bibfnamefont {C.}~\bibnamefont
  {McCabe}},\ }\href {\doibase 10.1103/PhysRevD.99.023012} {\bibfield
  {journal} {\bibinfo  {journal} {Phys. Rev. D}\ }\textbf {\bibinfo {volume}
  {99}},\ \bibinfo {pages} {023012} (\bibinfo {year} {2019})}\BibitemShut
  {NoStop}%
\bibitem [{\citenamefont {Lewin}\ and\ \citenamefont
  {Smith}(1996)}]{Lewin:1995rx}%
  \BibitemOpen
  \bibfield  {author} {\bibinfo {author} {\bibfnamefont {J.~D.}\ \bibnamefont
  {Lewin}}\ and\ \bibinfo {author} {\bibfnamefont {P.~F.}\ \bibnamefont
  {Smith}},\ }\href {\doibase 10.1016/S0927-6505(96)00047-3} {\bibfield
  {journal} {\bibinfo  {journal} {Astropart. Phys.}\ }\textbf {\bibinfo
  {volume} {6}},\ \bibinfo {pages} {87} (\bibinfo {year} {1996})}\BibitemShut
  {NoStop}%
\bibitem [{\citenamefont {Adhikari}\ \emph {et~al.}(2020)\citenamefont
  {Adhikari} \emph {et~al.}}]{DEAP:2020iwi}%
  \BibitemOpen
  \bibfield  {author} {\bibinfo {author} {\bibfnamefont {P.}~\bibnamefont
  {Adhikari}} \emph {et~al.} (\bibinfo {collaboration} {DEAP}),\ }\href
  {\doibase 10.1103/PhysRevD.102.082001} {\bibfield  {journal} {\bibinfo
  {journal} {Phys. Rev. D}\ }\textbf {\bibinfo {volume} {102}},\ \bibinfo
  {pages} {082001} (\bibinfo {year} {2020})},\ \bibinfo {note} {[Erratum:
  Phys.Rev.D 105, 029901 (2022)]},\ \Eprint {http://arxiv.org/abs/2005.14667}
  {arXiv:2005.14667 [astro-ph.CO]} \BibitemShut {NoStop}%
\bibitem [{\citenamefont {Smith-Orlik}\ \emph {et~al.}(2023)\citenamefont
  {Smith-Orlik} \emph {et~al.}}]{Smith-Orlik:2023kyl}%
  \BibitemOpen
  \bibfield  {author} {\bibinfo {author} {\bibfnamefont {A.}~\bibnamefont
  {Smith-Orlik}} \emph {et~al.},\ }\href {\doibase
  10.1088/1475-7516/2023/10/070} {\bibfield  {journal} {\bibinfo  {journal}
  {JCAP}\ }\textbf {\bibinfo {volume} {10}},\ \bibinfo {pages} {070} (\bibinfo
  {year} {2023})},\ \Eprint {http://arxiv.org/abs/2302.04281} {arXiv:2302.04281
  [astro-ph.GA]} \BibitemShut {NoStop}%
\bibitem [{\citenamefont {Gonzalo}\ and\ \citenamefont
  {Lucente}(2024)}]{Gonzalo:2023mdh}%
  \BibitemOpen
  \bibfield  {author} {\bibinfo {author} {\bibfnamefont {T.~E.}\ \bibnamefont
  {Gonzalo}}\ and\ \bibinfo {author} {\bibfnamefont {M.}~\bibnamefont
  {Lucente}},\ }\href {\doibase 10.1140/epjc/s10052-024-12423-3} {\bibfield
  {journal} {\bibinfo  {journal} {Eur. Phys. J. C}\ }\textbf {\bibinfo {volume}
  {84}},\ \bibinfo {pages} {119} (\bibinfo {year} {2024})},\ \Eprint
  {http://arxiv.org/abs/2303.15527} {arXiv:2303.15527 [hep-ph]} \BibitemShut
  {NoStop}%
\bibitem [{\citenamefont {Gonzalez-Garcia}\ \emph {et~al.}(2024)\citenamefont
  {Gonzalez-Garcia}, \citenamefont {Maltoni}, \citenamefont {Pinheiro},\ and\
  \citenamefont {Serenelli}}]{Gonzalez-Garcia:2023kva}%
  \BibitemOpen
  \bibfield  {author} {\bibinfo {author} {\bibfnamefont {M.~C.}\ \bibnamefont
  {Gonzalez-Garcia}}, \bibinfo {author} {\bibfnamefont {M.}~\bibnamefont
  {Maltoni}}, \bibinfo {author} {\bibfnamefont {J.~a.~P.}\ \bibnamefont
  {Pinheiro}}, \ and\ \bibinfo {author} {\bibfnamefont {A.~M.}\ \bibnamefont
  {Serenelli}},\ }\href {\doibase 10.1007/JHEP02(2024)064} {\bibfield
  {journal} {\bibinfo  {journal} {JHEP}\ }\textbf {\bibinfo {volume} {02}},\
  \bibinfo {pages} {064} (\bibinfo {year} {2024})},\ \Eprint
  {http://arxiv.org/abs/2311.16226} {arXiv:2311.16226 [hep-ph]} \BibitemShut
  {NoStop}%
\bibitem [{\citenamefont {Erler}\ and\ \citenamefont
  {Ramsey-Musolf}(2005)}]{erler2005weak}%
  \BibitemOpen
  \bibfield  {author} {\bibinfo {author} {\bibfnamefont {J.}~\bibnamefont
  {Erler}}\ and\ \bibinfo {author} {\bibfnamefont {M.~J.}\ \bibnamefont
  {Ramsey-Musolf}},\ }\href@noop {} {\bibfield  {journal} {\bibinfo  {journal}
  {Physical Review D}\ }\textbf {\bibinfo {volume} {72}},\ \bibinfo {pages}
  {073003} (\bibinfo {year} {2005})}\BibitemShut {NoStop}%
\bibitem [{\citenamefont {Barranco}\ \emph {et~al.}(2005)\citenamefont
  {Barranco}, \citenamefont {Miranda},\ and\ \citenamefont
  {Rashba}}]{barranco2005probing}%
  \BibitemOpen
  \bibfield  {author} {\bibinfo {author} {\bibfnamefont {J.}~\bibnamefont
  {Barranco}}, \bibinfo {author} {\bibfnamefont {O.~G.}\ \bibnamefont
  {Miranda}}, \ and\ \bibinfo {author} {\bibfnamefont {T.~I.}\ \bibnamefont
  {Rashba}},\ }\href@noop {} {\bibfield  {journal} {\bibinfo  {journal}
  {Journal of High Energy Physics}\ }\textbf {\bibinfo {volume} {2005}},\
  \bibinfo {pages} {021} (\bibinfo {year} {2005})}\BibitemShut {NoStop}%
\bibitem [{\citenamefont {Dutta}\ \emph {et~al.}(2016)\citenamefont {Dutta},
  \citenamefont {Gao}, \citenamefont {Kubik}, \citenamefont {Mahapatra},
  \citenamefont {Mirabolfathi}, \citenamefont {Strigari},\ and\ \citenamefont
  {Walker}}]{dutta2016sensitivity}%
  \BibitemOpen
  \bibfield  {author} {\bibinfo {author} {\bibfnamefont {B.}~\bibnamefont
  {Dutta}}, \bibinfo {author} {\bibfnamefont {Y.}~\bibnamefont {Gao}}, \bibinfo
  {author} {\bibfnamefont {A.}~\bibnamefont {Kubik}}, \bibinfo {author}
  {\bibfnamefont {R.}~\bibnamefont {Mahapatra}}, \bibinfo {author}
  {\bibfnamefont {N.}~\bibnamefont {Mirabolfathi}}, \bibinfo {author}
  {\bibfnamefont {L.~E.}\ \bibnamefont {Strigari}}, \ and\ \bibinfo {author}
  {\bibfnamefont {J.~W.}\ \bibnamefont {Walker}},\ }\href@noop {} {\bibfield
  {journal} {\bibinfo  {journal} {Physical Review D}\ }\textbf {\bibinfo
  {volume} {94}},\ \bibinfo {pages} {093002} (\bibinfo {year}
  {2016})}\BibitemShut {NoStop}%
\bibitem [{\citenamefont {Kosmas}\ and\ \citenamefont
  {Papoulias}(2015)}]{kosmas2015standard}%
  \BibitemOpen
  \bibfield  {author} {\bibinfo {author} {\bibfnamefont {T.}~\bibnamefont
  {Kosmas}}\ and\ \bibinfo {author} {\bibfnamefont {D.}~\bibnamefont
  {Papoulias}},\ }\href@noop {} {\bibfield  {journal} {\bibinfo  {journal}
  {Advances in High Energy Physics (Online)}\ }\textbf {\bibinfo {volume}
  {2015}} (\bibinfo {year} {2015})}\BibitemShut {NoStop}%
\bibitem [{\citenamefont {Scholberg}(2006)}]{scholberg2006prospects}%
  \BibitemOpen
  \bibfield  {author} {\bibinfo {author} {\bibfnamefont {K.}~\bibnamefont
  {Scholberg}},\ }\href@noop {} {\bibfield  {journal} {\bibinfo  {journal}
  {Physical Review D}\ }\textbf {\bibinfo {volume} {73}},\ \bibinfo {pages}
  {033005} (\bibinfo {year} {2006})}\BibitemShut {NoStop}%
\bibitem [{\citenamefont {Okada}(2018)}]{Okada:2018ktp}%
  \BibitemOpen
  \bibfield  {author} {\bibinfo {author} {\bibfnamefont {S.}~\bibnamefont
  {Okada}},\ }\href {\doibase 10.1155/2018/5340935} {\bibfield  {journal}
  {\bibinfo  {journal} {Adv. High Energy Phys.}\ }\textbf {\bibinfo {volume}
  {2018}},\ \bibinfo {pages} {5340935} (\bibinfo {year} {2018})},\ \Eprint
  {http://arxiv.org/abs/1803.06793} {arXiv:1803.06793 [hep-ph]} \BibitemShut
  {NoStop}%
\bibitem [{\citenamefont {Serenelli}\ \emph {et~al.}(2016)\citenamefont
  {Serenelli}, \citenamefont {Scott}, \citenamefont {Villante}, \citenamefont
  {Vincent}, \citenamefont {Asplund}, \citenamefont {Basu}, \citenamefont
  {Grevesse},\ and\ \citenamefont {Pena-Garay}}]{Serenelli:2016nms}%
  \BibitemOpen
  \bibfield  {author} {\bibinfo {author} {\bibfnamefont {A.}~\bibnamefont
  {Serenelli}}, \bibinfo {author} {\bibfnamefont {P.}~\bibnamefont {Scott}},
  \bibinfo {author} {\bibfnamefont {F.~L.}\ \bibnamefont {Villante}}, \bibinfo
  {author} {\bibfnamefont {A.~C.}\ \bibnamefont {Vincent}}, \bibinfo {author}
  {\bibfnamefont {M.}~\bibnamefont {Asplund}}, \bibinfo {author} {\bibfnamefont
  {S.}~\bibnamefont {Basu}}, \bibinfo {author} {\bibfnamefont {N.}~\bibnamefont
  {Grevesse}}, \ and\ \bibinfo {author} {\bibfnamefont {C.}~\bibnamefont
  {Pena-Garay}},\ }\href {\doibase 10.1093/mnras/stw1927} {\bibfield  {journal}
  {\bibinfo  {journal} {Mon. Not. Roy. Astron. Soc.}\ }\textbf {\bibinfo
  {volume} {463}},\ \bibinfo {pages} {2} (\bibinfo {year} {2016})},\ \Eprint
  {http://arxiv.org/abs/1604.05318} {arXiv:1604.05318 [astro-ph.SR]}
  \BibitemShut {NoStop}%
\bibitem [{\citenamefont {O'Hare}(2020)}]{OHare:2020lva}%
  \BibitemOpen
  \bibfield  {author} {\bibinfo {author} {\bibfnamefont {C.~A.~J.}\
  \bibnamefont {O'Hare}},\ }\href {\doibase 10.1103/PhysRevD.102.063024}
  {\bibfield  {journal} {\bibinfo  {journal} {Phys. Rev. D}\ }\textbf {\bibinfo
  {volume} {102}},\ \bibinfo {pages} {063024} (\bibinfo {year} {2020})},\
  \Eprint {http://arxiv.org/abs/2002.07499} {arXiv:2002.07499 [astro-ph.CO]}
  \BibitemShut {NoStop}%
\bibitem [{\citenamefont {Wilson}\ \emph {et~al.}(2009)\citenamefont {Wilson},
  \citenamefont {Perry}, \citenamefont {Charlton},\ and\ \citenamefont
  {Parish}}]{WilsonW.B.2009SAcf}%
  \BibitemOpen
  \bibfield  {author} {\bibinfo {author} {\bibfnamefont {W.}~\bibnamefont
  {Wilson}}, \bibinfo {author} {\bibfnamefont {R.}~\bibnamefont {Perry}},
  \bibinfo {author} {\bibfnamefont {W.}~\bibnamefont {Charlton}}, \ and\
  \bibinfo {author} {\bibfnamefont {T.}~\bibnamefont {Parish}},\ }\href@noop {}
  {\bibfield  {journal} {\bibinfo  {journal} {Progress in nuclear energy (New
  series)}\ }\textbf {\bibinfo {volume} {51}},\ \bibinfo {pages} {608}
  (\bibinfo {year} {2009})}\BibitemShut {NoStop}%
\bibitem [{\citenamefont {Koning}\ and\ \citenamefont
  {Rochman}(2012)}]{Koning:2012zqy}%
  \BibitemOpen
  \bibfield  {author} {\bibinfo {author} {\bibfnamefont {A.~J.}\ \bibnamefont
  {Koning}}\ and\ \bibinfo {author} {\bibfnamefont {D.}~\bibnamefont
  {Rochman}},\ }\href {\doibase 10.1016/j.nds.2012.11.002} {\bibfield
  {journal} {\bibinfo  {journal} {Nucl. Data Sheets}\ }\textbf {\bibinfo
  {volume} {113}},\ \bibinfo {pages} {2841} (\bibinfo {year}
  {2012})}\BibitemShut {NoStop}%
\bibitem [{\citenamefont {Plompen}\ \emph {et~al.}(2017)\citenamefont
  {Plompen}, \citenamefont {Hambsch}, \citenamefont {Schillebeeckx},
  \citenamefont {Mondelaers}, \citenamefont {Heyse}, \citenamefont {Kopecky},
  \citenamefont {Siegler},\ and\ \citenamefont {Oberstedt}}]{Plompen:2017lvx}%
  \BibitemOpen
  \bibinfo {editor} {\bibfnamefont {A.}~\bibnamefont {Plompen}}, \bibinfo
  {editor} {\bibfnamefont {F.~J.}\ \bibnamefont {Hambsch}}, \bibinfo {editor}
  {\bibfnamefont {P.}~\bibnamefont {Schillebeeckx}}, \bibinfo {editor}
  {\bibfnamefont {W.}~\bibnamefont {Mondelaers}}, \bibinfo {editor}
  {\bibfnamefont {J.}~\bibnamefont {Heyse}}, \bibinfo {editor} {\bibfnamefont
  {S.}~\bibnamefont {Kopecky}}, \bibinfo {editor} {\bibfnamefont
  {P.}~\bibnamefont {Siegler}}, \ and\ \bibinfo {editor} {\bibfnamefont
  {S.}~\bibnamefont {Oberstedt}},\ eds.,\ \href@noop {} {\emph {\bibinfo
  {title} {{Proceedings, ND 2016: International Conference on Nuclear Data for
  Science and Technology}: {Bruges, Belgium, September 11-16, 2016}}}}\
  (\bibinfo {year} {2017})\BibitemShut {NoStop}%
\bibitem [{\citenamefont {Sublet}\ \emph {et~al.}(2015)\citenamefont {Sublet},
  \citenamefont {Koning}, \citenamefont {Rochman}, \citenamefont {Fleming},\
  and\ \citenamefont {M.}}]{Sublet:2015}%
  \BibitemOpen
  \bibinfo {editor} {\bibfnamefont {J.~C.}\ \bibnamefont {Sublet}}, \bibinfo
  {editor} {\bibfnamefont {A.~J.}\ \bibnamefont {Koning}}, \bibinfo {editor}
  {\bibfnamefont {D.}~\bibnamefont {Rochman}}, \bibinfo {editor} {\bibfnamefont
  {M.}~\bibnamefont {Fleming}}, \ and\ \bibinfo {editor} {\bibfnamefont
  {G.}~\bibnamefont {M.}},\ eds.,\ \href@noop {} {\emph {\bibinfo {title}
  {{Proceedings, TENDL-2015: Delivering Both Completeness and Robustness,
  Advances in Nuclear Nonproliferation Technology and Policy Conference, Sept.
  25-30, Santa Fe, NM, USA}}}}\ (\bibinfo {year} {2015})\BibitemShut {NoStop}%
\bibitem [{\citenamefont {Fleming}\ \emph {et~al.}(2015)\citenamefont
  {Fleming}, \citenamefont {Sublet}, \citenamefont {Kopecky}, \citenamefont
  {Rochman},\ and\ \citenamefont {Koning}}]{Fleming:2015}%
  \BibitemOpen
  \bibfield  {author} {\bibinfo {author} {\bibfnamefont {M.}~\bibnamefont
  {Fleming}}, \bibinfo {author} {\bibfnamefont {J.~C.}\ \bibnamefont {Sublet}},
  \bibinfo {author} {\bibfnamefont {J.}~\bibnamefont {Kopecky}}, \bibinfo
  {author} {\bibfnamefont {D.}~\bibnamefont {Rochman}}, \ and\ \bibinfo
  {author} {\bibfnamefont {A.~J.}\ \bibnamefont {Koning}},\ }\href@noop {}
  {\bibfield  {journal} {\bibinfo  {journal} {CCFE report UKAEA-R(15)29}\ }
  (\bibinfo {year} {2015})}\BibitemShut {NoStop}%
\bibitem [{\citenamefont {Soppera}\ \emph {et~al.}(2014)\citenamefont
  {Soppera}, \citenamefont {Bossant},\ and\ \citenamefont
  {Dupont}}]{SopperaN.2014J4AI}%
  \BibitemOpen
  \bibfield  {author} {\bibinfo {author} {\bibfnamefont {N.}~\bibnamefont
  {Soppera}}, \bibinfo {author} {\bibfnamefont {M.}~\bibnamefont {Bossant}}, \
  and\ \bibinfo {author} {\bibfnamefont {E.}~\bibnamefont {Dupont}},\
  }\href@noop {} {\bibfield  {journal} {\bibinfo  {journal} {Nuclear data
  sheets}\ }\textbf {\bibinfo {volume} {120}},\ \bibinfo {pages} {294}
  (\bibinfo {year} {2014})}\BibitemShut {NoStop}%
\bibitem [{\citenamefont {Demouchy}(2021)}]{DemouchySylvie2021Dio}%
  \BibitemOpen
  \bibfield  {author} {\bibinfo {author} {\bibfnamefont {S.}~\bibnamefont
  {Demouchy}},\ }\href@noop {} {\bibfield  {journal} {\bibinfo  {journal}
  {European journal of mineralogy (Stuttgart)}\ }\textbf {\bibinfo {volume}
  {33}},\ \bibinfo {pages} {249} (\bibinfo {year} {2021})}\BibitemShut
  {NoStop}%
\bibitem [{\citenamefont {O'Hare}(2021)}]{OHare:2021utq}%
  \BibitemOpen
  \bibfield  {author} {\bibinfo {author} {\bibfnamefont {C.~A.~J.}\
  \bibnamefont {O'Hare}},\ }\href {\doibase 10.1103/PhysRevLett.127.251802}
  {\bibfield  {journal} {\bibinfo  {journal} {Phys. Rev. Lett.}\ }\textbf
  {\bibinfo {volume} {127}},\ \bibinfo {pages} {251802} (\bibinfo {year}
  {2021})},\ \Eprint {http://arxiv.org/abs/2109.03116} {arXiv:2109.03116
  [hep-ph]} \BibitemShut {NoStop}%
\bibitem [{\citenamefont {Aalbers}\ \emph
  {et~al.}(2024{\natexlab{b}})\citenamefont {Aalbers} \emph
  {et~al.}}]{LZ:2024zvo}%
  \BibitemOpen
  \bibfield  {author} {\bibinfo {author} {\bibfnamefont {J.}~\bibnamefont
  {Aalbers}} \emph {et~al.} (\bibinfo {collaboration} {LZ}),\ }\href@noop {} {\
   (\bibinfo {year} {2024}{\natexlab{b}})},\ \Eprint
  {http://arxiv.org/abs/2410.17036} {arXiv:2410.17036 [hep-ex]} \BibitemShut
  {NoStop}%
\bibitem [{\citenamefont {Agnes}\ \emph {et~al.}(2018)\citenamefont {Agnes}
  \emph {et~al.}}]{DarkSide:2018bpj}%
  \BibitemOpen
  \bibfield  {author} {\bibinfo {author} {\bibfnamefont {P.}~\bibnamefont
  {Agnes}} \emph {et~al.} (\bibinfo {collaboration} {DarkSide}),\ }\href
  {\doibase 10.1103/PhysRevLett.121.081307} {\bibfield  {journal} {\bibinfo
  {journal} {Phys. Rev. Lett.}\ }\textbf {\bibinfo {volume} {121}},\ \bibinfo
  {pages} {081307} (\bibinfo {year} {2018})},\ \Eprint
  {http://arxiv.org/abs/1802.06994} {arXiv:1802.06994 [astro-ph.HE]}
  \BibitemShut {NoStop}%
\bibitem [{\citenamefont {Abdelhameed}\ \emph {et~al.}(2019)\citenamefont
  {Abdelhameed} \emph {et~al.}}]{CRESST:2019jnq}%
  \BibitemOpen
  \bibfield  {author} {\bibinfo {author} {\bibfnamefont {A.~H.}\ \bibnamefont
  {Abdelhameed}} \emph {et~al.} (\bibinfo {collaboration} {CRESST}),\ }\href
  {\doibase 10.1103/PhysRevD.100.102002} {\bibfield  {journal} {\bibinfo
  {journal} {Phys. Rev. D}\ }\textbf {\bibinfo {volume} {100}},\ \bibinfo
  {pages} {102002} (\bibinfo {year} {2019})},\ \Eprint
  {http://arxiv.org/abs/1904.00498} {arXiv:1904.00498 [astro-ph.CO]}
  \BibitemShut {NoStop}%
\bibitem [{\citenamefont {Aprile}\ \emph {et~al.}(2019)\citenamefont {Aprile}
  \emph {et~al.}}]{XENON:2019gfn}%
  \BibitemOpen
  \bibfield  {author} {\bibinfo {author} {\bibfnamefont {E.}~\bibnamefont
  {Aprile}} \emph {et~al.} (\bibinfo {collaboration} {XENON}),\ }\href
  {\doibase 10.1103/PhysRevLett.123.251801} {\bibfield  {journal} {\bibinfo
  {journal} {Phys. Rev. Lett.}\ }\textbf {\bibinfo {volume} {123}},\ \bibinfo
  {pages} {251801} (\bibinfo {year} {2019})},\ \Eprint
  {http://arxiv.org/abs/1907.11485} {arXiv:1907.11485 [hep-ex]} \BibitemShut
  {NoStop}%
\bibitem [{\citenamefont {De~Romeri}\ \emph {et~al.}(2023)\citenamefont
  {De~Romeri}, \citenamefont {Miranda}, \citenamefont {Papoulias},
  \citenamefont {Sanchez~Garcia}, \citenamefont {T\'ortola},\ and\
  \citenamefont {Valle}}]{DeRomeri:2022twg}%
  \BibitemOpen
  \bibfield  {author} {\bibinfo {author} {\bibfnamefont {V.}~\bibnamefont
  {De~Romeri}}, \bibinfo {author} {\bibfnamefont {O.~G.}\ \bibnamefont
  {Miranda}}, \bibinfo {author} {\bibfnamefont {D.~K.}\ \bibnamefont
  {Papoulias}}, \bibinfo {author} {\bibfnamefont {G.}~\bibnamefont
  {Sanchez~Garcia}}, \bibinfo {author} {\bibfnamefont {M.}~\bibnamefont
  {T\'ortola}}, \ and\ \bibinfo {author} {\bibfnamefont {J.~W.~F.}\
  \bibnamefont {Valle}},\ }\href {\doibase 10.1007/JHEP04(2023)035} {\bibfield
  {journal} {\bibinfo  {journal} {JHEP}\ }\textbf {\bibinfo {volume} {04}},\
  \bibinfo {pages} {035} (\bibinfo {year} {2023})},\ \Eprint
  {http://arxiv.org/abs/2211.11905} {arXiv:2211.11905 [hep-ph]} \BibitemShut
  {NoStop}%
\bibitem [{\citenamefont {Atzori~Corona}\ \emph {et~al.}(2022)\citenamefont
  {Atzori~Corona}, \citenamefont {Cadeddu}, \citenamefont {Cargioli},
  \citenamefont {Dordei}, \citenamefont {Giunti}, \citenamefont {Li},
  \citenamefont {Picciau}, \citenamefont {Ternes},\ and\ \citenamefont
  {Zhang}}]{AtzoriCorona:2022moj}%
  \BibitemOpen
  \bibfield  {author} {\bibinfo {author} {\bibfnamefont {M.}~\bibnamefont
  {Atzori~Corona}}, \bibinfo {author} {\bibfnamefont {M.}~\bibnamefont
  {Cadeddu}}, \bibinfo {author} {\bibfnamefont {N.}~\bibnamefont {Cargioli}},
  \bibinfo {author} {\bibfnamefont {F.}~\bibnamefont {Dordei}}, \bibinfo
  {author} {\bibfnamefont {C.}~\bibnamefont {Giunti}}, \bibinfo {author}
  {\bibfnamefont {Y.~F.}\ \bibnamefont {Li}}, \bibinfo {author} {\bibfnamefont
  {E.}~\bibnamefont {Picciau}}, \bibinfo {author} {\bibfnamefont {C.~A.}\
  \bibnamefont {Ternes}}, \ and\ \bibinfo {author} {\bibfnamefont {Y.~Y.}\
  \bibnamefont {Zhang}},\ }\href {\doibase 10.1007/JHEP05(2022)109} {\bibfield
  {journal} {\bibinfo  {journal} {JHEP}\ }\textbf {\bibinfo {volume} {05}},\
  \bibinfo {pages} {109} (\bibinfo {year} {2022})},\ \Eprint
  {http://arxiv.org/abs/2202.11002} {arXiv:2202.11002 [hep-ph]} \BibitemShut
  {NoStop}%
\bibitem [{\citenamefont {Bonet}\ \emph {et~al.}(2022)\citenamefont {Bonet}
  \emph {et~al.}}]{CONUS:2021dwh}%
  \BibitemOpen
  \bibfield  {author} {\bibinfo {author} {\bibfnamefont {H.}~\bibnamefont
  {Bonet}} \emph {et~al.} (\bibinfo {collaboration} {CONUS}),\ }\href {\doibase
  10.1007/JHEP05(2022)085} {\bibfield  {journal} {\bibinfo  {journal} {JHEP}\
  }\textbf {\bibinfo {volume} {05}},\ \bibinfo {pages} {085} (\bibinfo {year}
  {2022})},\ \Eprint {http://arxiv.org/abs/2110.02174} {arXiv:2110.02174
  [hep-ph]} \BibitemShut {NoStop}%
\bibitem [{\citenamefont {Lindner}\ \emph {et~al.}(2024)\citenamefont
  {Lindner}, \citenamefont {Rink},\ and\ \citenamefont
  {Sen}}]{Lindner:2024eng}%
  \BibitemOpen
  \bibfield  {author} {\bibinfo {author} {\bibfnamefont {M.}~\bibnamefont
  {Lindner}}, \bibinfo {author} {\bibfnamefont {T.}~\bibnamefont {Rink}}, \
  and\ \bibinfo {author} {\bibfnamefont {M.}~\bibnamefont {Sen}},\ }\href
  {\doibase 10.1007/JHEP08(2024)171} {\bibfield  {journal} {\bibinfo  {journal}
  {JHEP}\ }\textbf {\bibinfo {volume} {08}},\ \bibinfo {pages} {171} (\bibinfo
  {year} {2024})},\ \Eprint {http://arxiv.org/abs/2401.13025} {arXiv:2401.13025
  [hep-ph]} \BibitemShut {NoStop}%
\bibitem [{\citenamefont {Aguilar-Arevalo}\ \emph {et~al.}(2020)\citenamefont
  {Aguilar-Arevalo} \emph {et~al.}}]{CONNIE:2019xid}%
  \BibitemOpen
  \bibfield  {author} {\bibinfo {author} {\bibfnamefont {A.}~\bibnamefont
  {Aguilar-Arevalo}} \emph {et~al.} (\bibinfo {collaboration} {CONNIE}),\
  }\href {\doibase 10.1007/JHEP04(2020)054} {\bibfield  {journal} {\bibinfo
  {journal} {JHEP}\ }\textbf {\bibinfo {volume} {04}},\ \bibinfo {pages} {054}
  (\bibinfo {year} {2020})},\ \Eprint {http://arxiv.org/abs/1910.04951}
  {arXiv:1910.04951 [hep-ex]} \BibitemShut {NoStop}%
\bibitem [{\citenamefont {Aguilar-Arevalo}\ \emph {et~al.}(2024)\citenamefont
  {Aguilar-Arevalo} \emph {et~al.}}]{CONNIE:2024pwt}%
  \BibitemOpen
  \bibfield  {author} {\bibinfo {author} {\bibfnamefont {A.~A.}\ \bibnamefont
  {Aguilar-Arevalo}} \emph {et~al.} (\bibinfo {collaboration} {CONNIE}),\
  }\href@noop {} {\  (\bibinfo {year} {2024})},\ \Eprint
  {http://arxiv.org/abs/2403.15976} {arXiv:2403.15976 [hep-ex]} \BibitemShut
  {NoStop}%
\bibitem [{\citenamefont {De~Romeri}\ \emph
  {et~al.}(2024{\natexlab{b}})\citenamefont {De~Romeri}, \citenamefont
  {Papoulias},\ and\ \citenamefont {Ternes}}]{DeRomeri:2024iaw}%
  \BibitemOpen
  \bibfield  {author} {\bibinfo {author} {\bibfnamefont {V.}~\bibnamefont
  {De~Romeri}}, \bibinfo {author} {\bibfnamefont {D.~K.}\ \bibnamefont
  {Papoulias}}, \ and\ \bibinfo {author} {\bibfnamefont {C.~A.}\ \bibnamefont
  {Ternes}},\ }\href@noop {} {\  (\bibinfo {year} {2024}{\natexlab{b}})},\
  \Eprint {http://arxiv.org/abs/2411.11749} {arXiv:2411.11749 [hep-ph]}
  \BibitemShut {NoStop}%
\bibitem [{\citenamefont {Bilmis}\ \emph {et~al.}(2015)\citenamefont {Bilmis},
  \citenamefont {Turan}, \citenamefont {Aliev}, \citenamefont {Deniz},
  \citenamefont {Singh},\ and\ \citenamefont {Wong}}]{Bilmis:2015lja}%
  \BibitemOpen
  \bibfield  {author} {\bibinfo {author} {\bibfnamefont {S.}~\bibnamefont
  {Bilmis}}, \bibinfo {author} {\bibfnamefont {I.}~\bibnamefont {Turan}},
  \bibinfo {author} {\bibfnamefont {T.~M.}\ \bibnamefont {Aliev}}, \bibinfo
  {author} {\bibfnamefont {M.}~\bibnamefont {Deniz}}, \bibinfo {author}
  {\bibfnamefont {L.}~\bibnamefont {Singh}}, \ and\ \bibinfo {author}
  {\bibfnamefont {H.~T.}\ \bibnamefont {Wong}},\ }\href {\doibase
  10.1103/PhysRevD.92.033009} {\bibfield  {journal} {\bibinfo  {journal} {Phys.
  Rev. D}\ }\textbf {\bibinfo {volume} {92}},\ \bibinfo {pages} {033009}
  (\bibinfo {year} {2015})},\ \Eprint {http://arxiv.org/abs/1502.07763}
  {arXiv:1502.07763 [hep-ph]} \BibitemShut {NoStop}%
\bibitem [{\citenamefont {Batell}\ \emph {et~al.}(2009)\citenamefont {Batell},
  \citenamefont {Pospelov},\ and\ \citenamefont {Ritz}}]{Batell:2009di}%
  \BibitemOpen
  \bibfield  {author} {\bibinfo {author} {\bibfnamefont {B.}~\bibnamefont
  {Batell}}, \bibinfo {author} {\bibfnamefont {M.}~\bibnamefont {Pospelov}}, \
  and\ \bibinfo {author} {\bibfnamefont {A.}~\bibnamefont {Ritz}},\ }\href
  {\doibase 10.1103/PhysRevD.80.095024} {\bibfield  {journal} {\bibinfo
  {journal} {Phys. Rev. D}\ }\textbf {\bibinfo {volume} {80}},\ \bibinfo
  {pages} {095024} (\bibinfo {year} {2009})},\ \Eprint
  {http://arxiv.org/abs/0906.5614} {arXiv:0906.5614 [hep-ph]} \BibitemShut
  {NoStop}%
\bibitem [{\citenamefont {Essig}\ \emph {et~al.}(2010)\citenamefont {Essig},
  \citenamefont {Harnik}, \citenamefont {Kaplan},\ and\ \citenamefont
  {Toro}}]{Essig:2010gu}%
  \BibitemOpen
  \bibfield  {author} {\bibinfo {author} {\bibfnamefont {R.}~\bibnamefont
  {Essig}}, \bibinfo {author} {\bibfnamefont {R.}~\bibnamefont {Harnik}},
  \bibinfo {author} {\bibfnamefont {J.}~\bibnamefont {Kaplan}}, \ and\ \bibinfo
  {author} {\bibfnamefont {N.}~\bibnamefont {Toro}},\ }\href {\doibase
  10.1103/PhysRevD.82.113008} {\bibfield  {journal} {\bibinfo  {journal} {Phys.
  Rev. D}\ }\textbf {\bibinfo {volume} {82}},\ \bibinfo {pages} {113008}
  (\bibinfo {year} {2010})},\ \Eprint {http://arxiv.org/abs/1008.0636}
  {arXiv:1008.0636 [hep-ph]} \BibitemShut {NoStop}%
\bibitem [{\citenamefont {Riordan}\ \emph {et~al.}(1987)\citenamefont {Riordan}
  \emph {et~al.}}]{Riordan:1987aw}%
  \BibitemOpen
  \bibfield  {author} {\bibinfo {author} {\bibfnamefont {E.~M.}\ \bibnamefont
  {Riordan}} \emph {et~al.},\ }\href {\doibase 10.1103/PhysRevLett.59.755}
  {\bibfield  {journal} {\bibinfo  {journal} {Phys. Rev. Lett.}\ }\textbf
  {\bibinfo {volume} {59}},\ \bibinfo {pages} {755} (\bibinfo {year}
  {1987})}\BibitemShut {NoStop}%
\bibitem [{\citenamefont {Bjorken}\ \emph {et~al.}(1988)\citenamefont
  {Bjorken}, \citenamefont {Ecklund}, \citenamefont {Nelson}, \citenamefont
  {Abashian}, \citenamefont {Church}, \citenamefont {Lu}, \citenamefont {Mo},
  \citenamefont {Nunamaker},\ and\ \citenamefont {Rassmann}}]{Bjorken:1988as}%
  \BibitemOpen
  \bibfield  {author} {\bibinfo {author} {\bibfnamefont {J.~D.}\ \bibnamefont
  {Bjorken}}, \bibinfo {author} {\bibfnamefont {S.}~\bibnamefont {Ecklund}},
  \bibinfo {author} {\bibfnamefont {W.~R.}\ \bibnamefont {Nelson}}, \bibinfo
  {author} {\bibfnamefont {A.}~\bibnamefont {Abashian}}, \bibinfo {author}
  {\bibfnamefont {C.}~\bibnamefont {Church}}, \bibinfo {author} {\bibfnamefont
  {B.}~\bibnamefont {Lu}}, \bibinfo {author} {\bibfnamefont {L.~W.}\
  \bibnamefont {Mo}}, \bibinfo {author} {\bibfnamefont {T.~A.}\ \bibnamefont
  {Nunamaker}}, \ and\ \bibinfo {author} {\bibfnamefont {P.}~\bibnamefont
  {Rassmann}},\ }\href {\doibase 10.1103/PhysRevD.38.3375} {\bibfield
  {journal} {\bibinfo  {journal} {Phys. Rev. D}\ }\textbf {\bibinfo {volume}
  {38}},\ \bibinfo {pages} {3375} (\bibinfo {year} {1988})}\BibitemShut
  {NoStop}%
\bibitem [{\citenamefont {Bross}\ \emph {et~al.}(1991)\citenamefont {Bross},
  \citenamefont {Crisler}, \citenamefont {Pordes}, \citenamefont {Volk},
  \citenamefont {Errede},\ and\ \citenamefont {Wrbanek}}]{Bross:1989mp}%
  \BibitemOpen
  \bibfield  {author} {\bibinfo {author} {\bibfnamefont {A.}~\bibnamefont
  {Bross}}, \bibinfo {author} {\bibfnamefont {M.}~\bibnamefont {Crisler}},
  \bibinfo {author} {\bibfnamefont {S.~H.}\ \bibnamefont {Pordes}}, \bibinfo
  {author} {\bibfnamefont {J.}~\bibnamefont {Volk}}, \bibinfo {author}
  {\bibfnamefont {S.}~\bibnamefont {Errede}}, \ and\ \bibinfo {author}
  {\bibfnamefont {J.}~\bibnamefont {Wrbanek}},\ }\href {\doibase
  10.1103/PhysRevLett.67.2942} {\bibfield  {journal} {\bibinfo  {journal}
  {Phys. Rev. Lett.}\ }\textbf {\bibinfo {volume} {67}},\ \bibinfo {pages}
  {2942} (\bibinfo {year} {1991})}\BibitemShut {NoStop}%
\bibitem [{\citenamefont {Konaka}\ \emph {et~al.}(1986)\citenamefont {Konaka}
  \emph {et~al.}}]{Konaka:1986cb}%
  \BibitemOpen
  \bibfield  {author} {\bibinfo {author} {\bibfnamefont {A.}~\bibnamefont
  {Konaka}} \emph {et~al.},\ }\href {\doibase 10.1103/PhysRevLett.57.659}
  {\bibfield  {journal} {\bibinfo  {journal} {Phys. Rev. Lett.}\ }\textbf
  {\bibinfo {volume} {57}},\ \bibinfo {pages} {659} (\bibinfo {year}
  {1986})}\BibitemShut {NoStop}%
\bibitem [{\citenamefont {Davier}\ and\ \citenamefont
  {Nguyen~Ngoc}(1989)}]{Davier:1989wz}%
  \BibitemOpen
  \bibfield  {author} {\bibinfo {author} {\bibfnamefont {M.}~\bibnamefont
  {Davier}}\ and\ \bibinfo {author} {\bibfnamefont {H.}~\bibnamefont
  {Nguyen~Ngoc}},\ }\href {\doibase 10.1016/0370-2693(89)90174-3} {\bibfield
  {journal} {\bibinfo  {journal} {Phys. Lett. B}\ }\textbf {\bibinfo {volume}
  {229}},\ \bibinfo {pages} {150} (\bibinfo {year} {1989})}\BibitemShut
  {NoStop}%
\bibitem [{\citenamefont {Bjorken}\ \emph {et~al.}(2009)\citenamefont
  {Bjorken}, \citenamefont {Essig}, \citenamefont {Schuster},\ and\
  \citenamefont {Toro}}]{Bjorken:2009mm}%
  \BibitemOpen
  \bibfield  {author} {\bibinfo {author} {\bibfnamefont {J.~D.}\ \bibnamefont
  {Bjorken}}, \bibinfo {author} {\bibfnamefont {R.}~\bibnamefont {Essig}},
  \bibinfo {author} {\bibfnamefont {P.}~\bibnamefont {Schuster}}, \ and\
  \bibinfo {author} {\bibfnamefont {N.}~\bibnamefont {Toro}},\ }\href {\doibase
  10.1103/PhysRevD.80.075018} {\bibfield  {journal} {\bibinfo  {journal} {Phys.
  Rev. D}\ }\textbf {\bibinfo {volume} {80}},\ \bibinfo {pages} {075018}
  (\bibinfo {year} {2009})},\ \Eprint {http://arxiv.org/abs/0906.0580}
  {arXiv:0906.0580 [hep-ph]} \BibitemShut {NoStop}%
\bibitem [{\citenamefont {Andreas}\ \emph {et~al.}(2012)\citenamefont
  {Andreas}, \citenamefont {Niebuhr},\ and\ \citenamefont
  {Ringwald}}]{Andreas:2012mt}%
  \BibitemOpen
  \bibfield  {author} {\bibinfo {author} {\bibfnamefont {S.}~\bibnamefont
  {Andreas}}, \bibinfo {author} {\bibfnamefont {C.}~\bibnamefont {Niebuhr}}, \
  and\ \bibinfo {author} {\bibfnamefont {A.}~\bibnamefont {Ringwald}},\ }\href
  {\doibase 10.1103/PhysRevD.86.095019} {\bibfield  {journal} {\bibinfo
  {journal} {Phys. Rev. D}\ }\textbf {\bibinfo {volume} {86}},\ \bibinfo
  {pages} {095019} (\bibinfo {year} {2012})},\ \Eprint
  {http://arxiv.org/abs/1209.6083} {arXiv:1209.6083 [hep-ph]} \BibitemShut
  {NoStop}%
\bibitem [{\citenamefont {Blumlein}\ \emph {et~al.}(1991)\citenamefont
  {Blumlein} \emph {et~al.}}]{Blumlein:1990ay}%
  \BibitemOpen
  \bibfield  {author} {\bibinfo {author} {\bibfnamefont {J.}~\bibnamefont
  {Blumlein}} \emph {et~al.},\ }\href {\doibase 10.1007/BF01548556} {\bibfield
  {journal} {\bibinfo  {journal} {Z. Phys. C}\ }\textbf {\bibinfo {volume}
  {51}},\ \bibinfo {pages} {341} (\bibinfo {year} {1991})}\BibitemShut
  {NoStop}%
\bibitem [{\citenamefont {Blumlein}\ \emph {et~al.}(1992)\citenamefont
  {Blumlein} \emph {et~al.}}]{Blumlein:1991xh}%
  \BibitemOpen
  \bibfield  {author} {\bibinfo {author} {\bibfnamefont {J.}~\bibnamefont
  {Blumlein}} \emph {et~al.},\ }\href {\doibase 10.1142/S0217751X9200171X}
  {\bibfield  {journal} {\bibinfo  {journal} {Int. J. Mod. Phys. A}\ }\textbf
  {\bibinfo {volume} {7}},\ \bibinfo {pages} {3835} (\bibinfo {year}
  {1992})}\BibitemShut {NoStop}%
\bibitem [{\citenamefont {Blumlein}\ and\ \citenamefont
  {Brunner}(2011)}]{Blumlein:2011mv}%
  \BibitemOpen
  \bibfield  {author} {\bibinfo {author} {\bibfnamefont {J.}~\bibnamefont
  {Blumlein}}\ and\ \bibinfo {author} {\bibfnamefont {J.}~\bibnamefont
  {Brunner}},\ }\href {\doibase 10.1016/j.physletb.2011.05.046} {\bibfield
  {journal} {\bibinfo  {journal} {Phys. Lett. B}\ }\textbf {\bibinfo {volume}
  {701}},\ \bibinfo {pages} {155} (\bibinfo {year} {2011})},\ \Eprint
  {http://arxiv.org/abs/1104.2747} {arXiv:1104.2747 [hep-ex]} \BibitemShut
  {NoStop}%
\bibitem [{\citenamefont {Bl\"umlein}\ and\ \citenamefont
  {Brunner}(2014)}]{Blumlein:2013cua}%
  \BibitemOpen
  \bibfield  {author} {\bibinfo {author} {\bibfnamefont {J.}~\bibnamefont
  {Bl\"umlein}}\ and\ \bibinfo {author} {\bibfnamefont {J.}~\bibnamefont
  {Brunner}},\ }\href {\doibase 10.1016/j.physletb.2014.02.029} {\bibfield
  {journal} {\bibinfo  {journal} {Phys. Lett. B}\ }\textbf {\bibinfo {volume}
  {731}},\ \bibinfo {pages} {320} (\bibinfo {year} {2014})},\ \Eprint
  {http://arxiv.org/abs/1311.3870} {arXiv:1311.3870 [hep-ph]} \BibitemShut
  {NoStop}%
\bibitem [{\citenamefont {Bergsma}\ \emph {et~al.}(1985)\citenamefont {Bergsma}
  \emph {et~al.}}]{CHARM:1985anb}%
  \BibitemOpen
  \bibfield  {author} {\bibinfo {author} {\bibfnamefont {F.}~\bibnamefont
  {Bergsma}} \emph {et~al.} (\bibinfo {collaboration} {CHARM}),\ }\href
  {\doibase 10.1016/0370-2693(85)90400-9} {\bibfield  {journal} {\bibinfo
  {journal} {Phys. Lett. B}\ }\textbf {\bibinfo {volume} {157}},\ \bibinfo
  {pages} {458} (\bibinfo {year} {1985})}\BibitemShut {NoStop}%
\bibitem [{\citenamefont {Gninenko}(2012{\natexlab{a}})}]{Gninenko:2012eq}%
  \BibitemOpen
  \bibfield  {author} {\bibinfo {author} {\bibfnamefont {S.~N.}\ \bibnamefont
  {Gninenko}},\ }\href {\doibase 10.1016/j.physletb.2012.06.002} {\bibfield
  {journal} {\bibinfo  {journal} {Phys. Lett. B}\ }\textbf {\bibinfo {volume}
  {713}},\ \bibinfo {pages} {244} (\bibinfo {year} {2012}{\natexlab{a}})},\
  \Eprint {http://arxiv.org/abs/1204.3583} {arXiv:1204.3583 [hep-ph]}
  \BibitemShut {NoStop}%
\bibitem [{\citenamefont {Astier}\ \emph {et~al.}(2001)\citenamefont {Astier}
  \emph {et~al.}}]{NOMAD:2001eyx}%
  \BibitemOpen
  \bibfield  {author} {\bibinfo {author} {\bibfnamefont {P.}~\bibnamefont
  {Astier}} \emph {et~al.} (\bibinfo {collaboration} {NOMAD}),\ }\href
  {\doibase 10.1016/S0370-2693(01)00362-8} {\bibfield  {journal} {\bibinfo
  {journal} {Phys. Lett. B}\ }\textbf {\bibinfo {volume} {506}},\ \bibinfo
  {pages} {27} (\bibinfo {year} {2001})},\ \Eprint
  {http://arxiv.org/abs/hep-ex/0101041} {arXiv:hep-ex/0101041} \BibitemShut
  {NoStop}%
\bibitem [{\citenamefont {Bernardi}\ \emph {et~al.}(1986)\citenamefont
  {Bernardi} \emph {et~al.}}]{Bernardi:1985ny}%
  \BibitemOpen
  \bibfield  {author} {\bibinfo {author} {\bibfnamefont {G.}~\bibnamefont
  {Bernardi}} \emph {et~al.},\ }\href {\doibase 10.1016/0370-2693(86)91602-3}
  {\bibfield  {journal} {\bibinfo  {journal} {Phys. Lett. B}\ }\textbf
  {\bibinfo {volume} {166}},\ \bibinfo {pages} {479} (\bibinfo {year}
  {1986})}\BibitemShut {NoStop}%
\bibitem [{\citenamefont {Gninenko}(2012{\natexlab{b}})}]{Gninenko:2011uv}%
  \BibitemOpen
  \bibfield  {author} {\bibinfo {author} {\bibfnamefont {S.~N.}\ \bibnamefont
  {Gninenko}},\ }\href {\doibase 10.1103/PhysRevD.85.055027} {\bibfield
  {journal} {\bibinfo  {journal} {Phys. Rev. D}\ }\textbf {\bibinfo {volume}
  {85}},\ \bibinfo {pages} {055027} (\bibinfo {year} {2012}{\natexlab{b}})},\
  \Eprint {http://arxiv.org/abs/1112.5438} {arXiv:1112.5438 [hep-ph]}
  \BibitemShut {NoStop}%
\bibitem [{\citenamefont {Merkel}\ \emph {et~al.}(2014)\citenamefont {Merkel}
  \emph {et~al.}}]{Merkel:2014avp}%
  \BibitemOpen
  \bibfield  {author} {\bibinfo {author} {\bibfnamefont {H.}~\bibnamefont
  {Merkel}} \emph {et~al.},\ }\href {\doibase 10.1103/PhysRevLett.112.221802}
  {\bibfield  {journal} {\bibinfo  {journal} {Phys. Rev. Lett.}\ }\textbf
  {\bibinfo {volume} {112}},\ \bibinfo {pages} {221802} (\bibinfo {year}
  {2014})},\ \Eprint {http://arxiv.org/abs/1404.5502} {arXiv:1404.5502
  [hep-ex]} \BibitemShut {NoStop}%
\bibitem [{\citenamefont {Abrahamyan}\ \emph {et~al.}(2011)\citenamefont
  {Abrahamyan} \emph {et~al.}}]{APEX:2011dww}%
  \BibitemOpen
  \bibfield  {author} {\bibinfo {author} {\bibfnamefont {S.}~\bibnamefont
  {Abrahamyan}} \emph {et~al.} (\bibinfo {collaboration} {APEX}),\ }\href
  {\doibase 10.1103/PhysRevLett.107.191804} {\bibfield  {journal} {\bibinfo
  {journal} {Phys. Rev. Lett.}\ }\textbf {\bibinfo {volume} {107}},\ \bibinfo
  {pages} {191804} (\bibinfo {year} {2011})},\ \Eprint
  {http://arxiv.org/abs/1108.2750} {arXiv:1108.2750 [hep-ex]} \BibitemShut
  {NoStop}%
\bibitem [{\citenamefont {Rrapaj}\ and\ \citenamefont
  {Reddy}(2016)}]{Rrapaj:2015wgs}%
  \BibitemOpen
  \bibfield  {author} {\bibinfo {author} {\bibfnamefont {E.}~\bibnamefont
  {Rrapaj}}\ and\ \bibinfo {author} {\bibfnamefont {S.}~\bibnamefont {Reddy}},\
  }\href {\doibase 10.1103/PhysRevC.94.045805} {\bibfield  {journal} {\bibinfo
  {journal} {Phys. Rev. C}\ }\textbf {\bibinfo {volume} {94}},\ \bibinfo
  {pages} {045805} (\bibinfo {year} {2016})},\ \Eprint
  {http://arxiv.org/abs/1511.09136} {arXiv:1511.09136 [nucl-th]} \BibitemShut
  {NoStop}%
\bibitem [{\citenamefont {Harnik}\ \emph {et~al.}(2012)\citenamefont {Harnik},
  \citenamefont {Kopp},\ and\ \citenamefont {Machado}}]{Harnik:2012ni}%
  \BibitemOpen
  \bibfield  {author} {\bibinfo {author} {\bibfnamefont {R.}~\bibnamefont
  {Harnik}}, \bibinfo {author} {\bibfnamefont {J.}~\bibnamefont {Kopp}}, \ and\
  \bibinfo {author} {\bibfnamefont {P.~A.~N.}\ \bibnamefont {Machado}},\ }\href
  {\doibase 10.1088/1475-7516/2012/07/026} {\bibfield  {journal} {\bibinfo
  {journal} {JCAP}\ }\textbf {\bibinfo {volume} {07}},\ \bibinfo {pages} {026}
  (\bibinfo {year} {2012})},\ \Eprint {http://arxiv.org/abs/1202.6073}
  {arXiv:1202.6073 [hep-ph]} \BibitemShut {NoStop}%
\bibitem [{\citenamefont {Ghosh}\ \emph {et~al.}(2024)\citenamefont {Ghosh},
  \citenamefont {Ghosh}, \citenamefont {Jeesun},\ and\ \citenamefont
  {Srivastava}}]{Ghosh:2024cxi}%
  \BibitemOpen
  \bibfield  {author} {\bibinfo {author} {\bibfnamefont {D.~K.}\ \bibnamefont
  {Ghosh}}, \bibinfo {author} {\bibfnamefont {P.}~\bibnamefont {Ghosh}},
  \bibinfo {author} {\bibfnamefont {S.}~\bibnamefont {Jeesun}}, \ and\ \bibinfo
  {author} {\bibfnamefont {R.}~\bibnamefont {Srivastava}},\ }\href {\doibase
  10.1103/PhysRevD.110.075032} {\bibfield  {journal} {\bibinfo  {journal}
  {Phys. Rev. D}\ }\textbf {\bibinfo {volume} {110}},\ \bibinfo {pages}
  {075032} (\bibinfo {year} {2024})},\ \Eprint
  {http://arxiv.org/abs/2404.10077} {arXiv:2404.10077 [hep-ph]} \BibitemShut
  {NoStop}%
\bibitem [{\citenamefont {Redondo}(2008)}]{Redondo:2008aa}%
  \BibitemOpen
  \bibfield  {author} {\bibinfo {author} {\bibfnamefont {J.}~\bibnamefont
  {Redondo}},\ }\href {\doibase 10.1088/1475-7516/2008/07/008} {\bibfield
  {journal} {\bibinfo  {journal} {JCAP}\ }\textbf {\bibinfo {volume} {07}},\
  \bibinfo {pages} {008} (\bibinfo {year} {2008})},\ \Eprint
  {http://arxiv.org/abs/0801.1527} {arXiv:0801.1527 [hep-ph]} \BibitemShut
  {NoStop}%
\end{thebibliography}%

\appendix
\subsection{Bézier parametrisation of tracks}
\label{sec: bezier}
Bézier curves $\mathbf{B}(t)$, with parameter $0\leq t \leq 1$, are defined by the positions of control points $\mathbf{P}_i$ that are fixed in space. It is the locus of a line segment which endpoints are constantly changing as $t$ increases, whose locations traces the loci of each line segment formed by each consecutive pair of control points $\mathbf{P}_i$ and $\mathbf{P}_{i+1}$. Mathematically, a Bézier curve with degree $n$ is defined as
\begin{equation}
    \mathbf{B}(t) \equiv \sum_{k=0}^{n} b^k_n(t) \mathbf{P}_k
    \label{eq:generalBezier}
\end{equation}
where $\mathbf{B}(t)$ and $\mathbf{P}_k$ are 3-vectors in 3D space, the subscript $k$ represents the $k-th$ control point, $n+1$ is the total number of control points, and $b^k_n(t)$ is the Bernstein polynomial
\begin{equation}
    b^k_n(t) = \binom{n}{k}t^k(1-t)^{n-k}. 
\end{equation}
Equation \ref{eq:generalBezier} can also be written in matrix form:
\begin{gather}
\label{eq:bezier_matrix_eq}
 \begin{bmatrix} B_x(t) \\ B_y(t) \\ B_z(t)  \end{bmatrix}
 =
  \begin{bmatrix}
   P_{x,0} & P_{x,1} & ...\ P_{x,n}\\ 
   P_{y,0} & P_{y,1} & ...\ P_{y,n}\\
   P_{z,0} & P_{z,1} & ...\ P_{z,n}\\
   \end{bmatrix}
\begin{bmatrix} b_{0,n}(t) \\ b_{1,n}(t) \\ . \\ . \\ . \\ b_{n,n}(t)
\end{bmatrix},
\end{gather}
for Bézier curves in 3D space with $n+1$ many control points. 
In the context of parameterizing track lengths, the coordinates of the displaced atoms obtained from \trim simulations, shown as scatter dots in Figure \ref{fig:bezier_track}, are taken to be $\mathbf{B}(t)$, where $t=0$ corresponds to the first recoil (the leftmost red dot in Figure. \ref{fig:bezier_track}). By inverting Eq. \eqref{eq:bezier_matrix_eq}, we obtain the coordinates of $n$ control points $\mathbf{P}_k$ necessary to reconstruct a parametric curve that passes through all vacant lattice sites with degree $n$. We increase the degree $n$ from $0$ until the least-square error is just above the chosen resolution. 

\end{document}